\documentclass[lettersize,journal]{IEEEtran}
\usepackage[switch]{lineno}

\usepackage{array}
\usepackage{cite}
\usepackage{amsmath,amssymb,amsfonts}
\usepackage{algorithmic,algorithm}
\usepackage{graphicx,dblfloatfix}
\usepackage{textcomp}
\usepackage{multirow}
\usepackage{booktabs}
\usepackage{xcolor}
\usepackage{url}
\usepackage{enumitem}
\usepackage{threeparttable}

\newcommand{\bh}[1]{\vspace{3pt}\noindent \textbf{#1} }

\begin{document}

\title{
APEX: Adaptive Expert Prefetching for Memory-Efficient Edge MoE Inference
}

\author{%
\IEEEauthorblockN{%
Alish~Kanani,
Layan Badawi and
Umit Y.~Ogras}\\

\IEEEauthorblockA{%
University of Wisconsin--Madison;~
\{ahkanani, lbadawi, uogras\}@wisc.edu
\vspace{-7mm}}

\thanks{This work was partially supported by the Intel CAD SRS program.}}

\maketitle
\begin{abstract}
Mixture-of-Experts (MoE) models are attractive for edge deployment because they provide high model capacity while activating only a small subset of parameters per token, improving compute efficiency. 
However, MoE inference at the edge is fundamentally limited by memory. 
Expert parameters are large and often reside in off-chip memory due to capacity, cost, and power constraints, putting expert loading to the critical path. 
We present \emph{APEX: Adaptive Expert Prefetching}, a predictive resource management framework that overlaps expert loading with useful computation. 
APEX introduces a lightweight prefetch router that predicts candidate experts before the attention block to dynamically fetch additional experts using a learned confidence model. 
This adaptive strategy achieves over 99\% overlap accuracy, significantly outperforming fixed top-$k$ prefetching techniques. 
APEX supports two execution modes: a correctness-preserving mode that guarantees exact routing semantics, and a stall-free mode that eliminates residual stalls by operating on available experts with negligible impact on application accuracy. 
Across multiple MoE models, the correctness-preserving mode reduces per-token latency by up to 26\% and improves energy-delay product (EDP) by up to 41\% over state-of-the-art baselines, while the stall-free mode provides additional efficiency gains with negligible impact on application accuracy.
These results establish adaptive, confidence-driven expert prefetching as an effective approach for efficient MoE inference on edge systems.
\end{abstract}

%

\section{Introduction}\label{sec:intro}

Large language models (LLMs) are increasingly deployed at the edge, where low-latency response, energy efficiency, data privacy, and reduced cloud dependence are critical requirements~\cite{meuser2024revisiting,zheng2025review}. 
Mixture-of-Experts (MoE) models are particularly well-suited for this setting, as they decouple model capacity from per-token compute~\cite{fedus2022switch,jiang2024mixtral}. 
By activating only a small subset of experts per token, MoEs achieve high accuracy while operating within tight compute (FLOPs) budgets. 
This technique enables edge accelerators to support much larger models that would otherwise be computationally prohibitive.

MoE inference at the edge faces a fundamental memory bottleneck~\cite{gholami2024ai}. 
Unlike dense models with deterministic weights, MoEs repeatedly fetch large, sparse, and irregularly accessed expert parameters~\cite{lepikhin2020gshard,pope2023efficiently}. 
Keeping all experts resident in high-bandwidth on-package memory (e.g., HBM or GDDR) is impractical on edge platforms due to capacity, cost, thermal, and power constraints~\cite{zheng2025review,park2024thermal,pfromm2025mfit}. 
In practice, expert parameters are stored in lower-cost off-chip memory (e.g., LPDDR), while non-expert weights, activations, and the key-value (KV) cache remain on-package. 
Although this organization is more realistic for edge AI accelerators, it places expert loading on the critical path of token generation. 

\begin{figure}[t]
    \centering
    \includegraphics[width=0.9\linewidth]{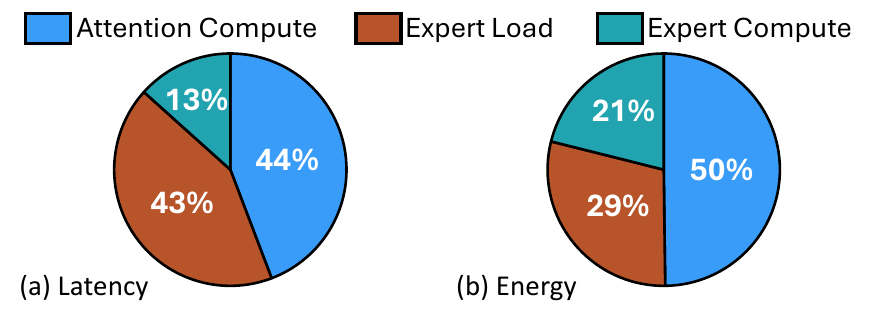}
    \vspace{-3mm}
    \caption{Latency and energy breakdown for the Granite-3.1-3B-A800M model on a representative edge platform with off-chip expert memory and 1024-token context length.} 
    \vspace{-3mm}
    \label{fig:motivation}
\end{figure}

Figure~\ref{fig:motivation} illustrates the memory bottleneck on a representative high-end edge accelerator composed of a compute chiplet with on-package memory and an I/O chiplet. The modeled compute capability and on-package memory bandwidth are comparable to high-end edge GPU platforms such as NVIDIA RTX 3090~\cite{NVIDIA_RTX3090}.
To reflect practical deployments, expert weights reside in host-attached LPDDR5X and are streamed to the accelerator over a PCIe~6.0 $\times$16 link. Non-expert weights, activations, and the KV cache remain on-package. Full timing and per-bit energy parameters are provided in Section~\ref{sec:eval}.
Under this realistic memory hierarchy, expert loading becomes a dominant cost.
For single-token generation with the Granite-3.1-3B-A800M model~\cite{granite31_a800m}, expert loading contributes 43\% of latency and 29\% of total energy. 
Importantly, this cost is not purely a bandwidth problem. 
The compute package remains underutilized yet continues to dissipate static power while waiting for expert transfers.

We argue that the solution is not static memory scaling, but \emph{adaptive resource management}~\cite{kanani2025thermos}.
If the required experts can be predicted early, they can be prefetched to reduce exposed I/O latency. 
Prior work follows this idea by \textit{statically} prefetching top-$k$ predicted experts, assuming that prediction accuracy (i.e., the overlap between the predicted and actual top-$k$ experts) is sufficient to ensure coverage~\cite{song2024promoe}.
However, routing uncertainty varies significantly across layers and tokens, and prediction errors can lead to missing experts. 
Even a relatively high \textit{average} prediction accuracy (e.g., 70--85\%) is insufficient, since some layers may miss consistently more experts and experience stalls that negate much of the prefetching benefit~\cite{song2024promoe,zhong2024adapmoe,tang2024hobbit}. 
Hence, existing approaches with a fixed prefetch budget fail to adapt to layer and input variations, leaving significant performance and energy gains unrealized.

\begin{table*}[b]
\vspace{-3mm}
\caption{Comparison of state-of-the-art MoE inference methods for edge deployment}
\vspace{-2mm}
\label{tab:comparison}
\centering
\begin{threeparttable}
\resizebox{\linewidth}{!}{%
\begin{tabular}{@{}lccccc@{}}
\toprule
Approach & \textbf{Early Expert Prediction} & \textbf{Dynamic Prefetch Budget} & \textbf{Correctness Guarantee} & \textbf{Stall-free} & \textbf{Key Idea} \\ \midrule
Pre-gated MoE~\cite{hwang2024pre} & $\times$ & $\times$ & $\times$ & $\checkmark$ &  Router in layer $L$ selects experts for layer $L+1$ \\
MoE-Infinity~\cite{xue2024moe} & $\times$ & $\times$ & $\checkmark$ & $\times$ & Prefetching is guided by historical traces\\
Fate~\cite{fang2025fate} & $\checkmark$ & $\times$ & $\times$ & $\times$ & Cross-Layer prefetch router \\
HOBBIT~\cite{tang2024hobbit} & $\checkmark$ & $\times$ \tnote{a} & $\times$ & $\times$ & Prefetching mixed precision experts  \\
Pre-Attn Pred. ~\cite{zhu2025pre} &  $\checkmark$ &  $\times$ & $\checkmark$ & $\times$ & Same-layer pre-attention expert prediction \\
ProMoE~\cite{song2024promoe} & $\checkmark$ & $\times$ & $\checkmark$ & $\times$ & Learned proactive expert caching \\
\textbf{APEX} & $\pmb\checkmark$ & $\pmb\checkmark$ & $\pmb\checkmark$ & $\pmb\checkmark$ & 
\textbf{\begin{tabular}[c]{@{}c@{}} Adaptive expert prefetching \\ using per-token confidence\end{tabular}}\\ \bottomrule
\end{tabular}
}
\begin{tablenotes}
\footnotesize
\item[a] It uses an adaptive predictor that continues predicting future layers while the predicted experts remain in cache, but it is still not adaptive per-token.
\end{tablenotes}
\end{threeparttable}
\end{table*}

We address this problem with \emph{APEX, an adaptive expert prefetching} framework that overlaps expert loading with attention computation. 
APEX adds a lightweight prefetch router before the attention block to predict a ranked list of candidate experts for the current layer. 
Rather than always prefetching a fixed top-$k$ set, APEX exploits the observation that most routing misses can be avoided by fetching only a small, token-dependent number of additional experts. 
Therefore, it prefetches a top-$(k+\hat{\delta}(x))$ set, where $\hat{\delta}(x)$ is \textit{dynamically selected per token} using a learned confidence model. 
This enables APEX to fetch \emph{just enough} additional experts to suppress misses without paying the energy cost of aggressive over-prefetching on every token.
While fetching additional experts slightly increases communication energy, it substantially reduces the latency and stall time associated with exposed expert loading, resulting in a net gain in both energy and performance. 
APEX supports two execution modes: (1)~\textbf{a correctness-preserving mode} guarantees exact routing semantics by fetching any rarely missing experts, and (2)~\textbf{an optional stall-free mode}, which eliminates residual stalls by operating on available experts when the application can tolerate the resulting accuracy trade-off.

We evaluate APEX across IBM Granite-3.1-1B-A400M~\cite{granite31_1b_a400m}, Granite-3.1-3B-A800M~\cite{granite31_a800m}, Microsoft Phi-mini-MoE-7B-A2.4B~\cite{phi3_technical_report} and DeepSeek-V2-Lite-16B-A2.4B~\cite{liu2024deepseekv2}
models on edge configurations with off-chip expert memory. 
It achieves near-oracle overlap ($>$99\%), substantially reducing expert-loading stalls and improving end-to-end EDP by up to 41\% over state-of-the-art baselines.
We further show that the stall-free variant achieves an additional 2--14\% EDP improvement with negligible impact on application accuracy.
These results demonstrate that adaptive expert prefetching is a practical and effective approach for enabling efficient MoE inference at the edge.

The key contributions of this work are as follows:
\begin{itemize}[leftmargin=*]
    \item An adaptive top-($k+\hat\delta(x)$) prefetching strategy that adaptively selects the number of prefetched experts using a learned confidence model. 
    
    \item Two execution modes: a correctness-preserving mode, which  guarantees exact routing semantics, and an optional stall-free mode, which further reduces latency and energy with negligible impact on application accuracy.
    
    \item Comprehensive evaluations across multiple MoE models on an edge platform with off-chip expert memory, achieving $>$99\% expert overlap and up to 41\% improvement in end-to-end EDP over state-of-the-art baselines.

\end{itemize}

In the rest, Section~\ref{sec:related} reviews related work, while Section~\ref{sec:background} provides background on MoE inference.
Section~\ref{sec:methodology} presents our adaptive expert prefetching method, followed by evaluations in Section~\ref{sec:eval}, and Section~\ref{sec:conclusion} concludes the paper.

\section{Related Work}\label{sec:related}

MoE models scale model capacity by replacing dense feed-forward networks with multiple experts and a router that activates only a sparse subset per token~\cite{shazeer2017outrageously}. 
This conditional-computation paradigm has been extended to large-scale training systems such as GShard~\cite{lepikhin2020gshard} and later improved through routing stability, load-balancing, and training-stability techniques~\cite{zoph2022st,lewis2021base,zou2024fed,cheng2025ermoe}.
More recently, MoE architectures have been adopted in both open and proprietary LLMs. 
Models such as Mixtral~\cite{jiang2024mixtral}, DeepSeekMoE~\cite{dai2024deepseekmoe}, and OLMoE~\cite{muennighoff2024olmoe} extend MoE designs to modern decoder-only LLMs, enabling efficient scaling in practical deployments.

A substantial body of work addresses system-level challenges in MoE training and inference at a datacenter scale~\cite{he2025hydra,dong2025ubimoe}. 
Early systems, such as FastMoE~\cite{he2021fastmoe}, DeepSpeed-MoE~\cite{aminabadi2022deepspeed}, and MegaBlocks~\cite{gale2023megablocks}, optimize token dispatch, expert parallelism, and all-to-all communication to enable efficient large-scale training. 
However, these systems assume that expert parameters reside in accelerator memory and rely on high-bandwidth interconnects. 
These assumptions do not hold in edge deployments, where memory capacity is limited, and experts are often offloaded to external memory. 
As a result, expert movement becomes a dominant runtime bottleneck.

Hot-expert caching methods, such as MoE-Infinity~\cite{xue2024moe}, exploit temporal locality by keeping frequently activated experts in GPU memory and prefetching others from host memory. 
This is effective when persistent device memory is available and expert reuse is stable across tokens or requests. 
In edge settings, however, on-package memory must also hold non-expert weights, activations, the KV cache, and current-layer staging buffers. 
Reserving capacity for static hot experts can reduce the space available for token-specific expert movement. 

MoE optimization methods for edge inference can be broadly divided into heuristic-based and predictive approaches, as summarized in Table~\ref{tab:comparison}. 
Heuristic methods reduce memory pressure through caching, reuse, or architectural modifications, without explicitly predicting future expert usage. 
Pre-gated MoE shifts routing earlier by having layer $L$ select experts for layer $L+1$, enabling partial overlap through buffering, but cannot preserve the correctness of the original routing decisions~\cite{hwang2024pre}.
AdapMoE combines cross-layer prefetching and cache allocation, but changes the number of active experts and is orthogonal to fixed top-$k$ routing~\cite{zhong2024adapmoe}. 
MoE-SpAc uses speculative decoding to anticipate expert demand, but it operates at the sequence level rather than modeling per-layer routing decisions~\cite{li2026moe}.

Predictive approaches forecast future expert usage and prefetch weights ahead of execution to hide data movement latency. 
Fate~\cite{fang2025fate} leverages cross-layer routing similarity to predict expert usage in subsequent layers, while 
HOBBIT~\cite{tang2024hobbit} extends this idea with mixed-precision expert loading, combining prediction with hardware-aware optimizations, but without strict correctness guarantees. 
Recent pre-attention expert prediction work~\cite{zhu2025pre} uses same-layer pre-attention activations to predict expert selections earlier in the layer, improving prediction accuracy.
While such methods improve the predictor, they still rely on a fixed predicted set unlike APEX, which focuses on adaptive resource management.
ProMoE, the closest state-of-the-art work, uses a learned predictor based on intermediate activations to anticipate routing decisions and prefetch experts~\cite{song2024promoe}.  
However, it employs a static top-$k$ prefetch policy that does not adapt the prefetch budget, which we evaluate as a primary baseline in Section~\ref{sec:eval}.

Despite these advances, existing prediction-based methods prefetch only the predicted top-$k$ experts. 
As shown in Table~\ref{tab:comparison}, none of them adapts the prefetch budget using prediction confidence. 
In contrast, APEX uses an adaptive top-$(k+\hat{\delta}(x))$ prefetch strategy that dynamically determines the minimal additional budget needed to meet a target coverage. 
By using a learned confidence model, APEX improves coverage reliability while maintaining energy efficiency and enabling both correctness-preserving and stall-free execution modes.

\section{Background and Motivation}\label{sec:background}

This section reviews the execution characteristics of MoE layers in edge settings, highlights the resulting memory bottlenecks, and discusses the limitations of static prefetching.

\subsection{MoE Inference at the Edge}\label{ssec:basics}

\begin{figure}[t]
    \centering
    \includegraphics[width=\linewidth]{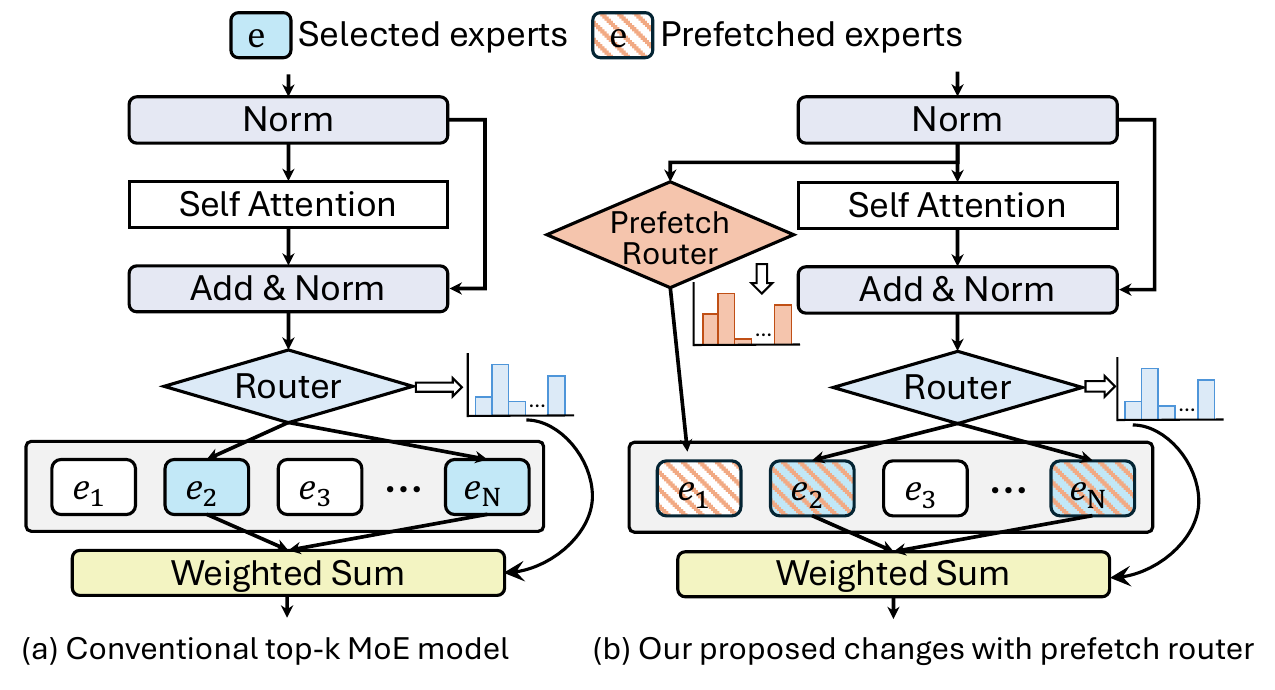}
    \caption{Comparison of (a) standard MoE block; and (b) our proposed architecture with an additional prefetch router.}
    \label{fig:moe_arch}
\end{figure}

Modern LLMs are built on the Transformer architecture, where each layer consists of a self-attention block and a feed-forward network (FFN)~\cite{vaswani2017attention}. 
The attention mechanism captures token-to-token interactions for contextual reasoning, while the FFN provides most of the model capacity.
In MoE models, the dense FFN is replaced by a set of $N$ experts, each implemented as a smaller FFN and a router, as shown in Figure~\ref{fig:moe_arch}(a). 
For each token, the router computes a probability distribution over experts and selects the top-$k$ experts~\cite{shazeer2017outrageously}. 
These experts process the token independently, and their outputs are combined using a weighted sum. 
Training objectives encourage balanced expert utilization~\cite{lepikhin2020gshard}.
Thus, routing decisions vary across tokens and layers, making expert selection prediction challenging at inference.

A key property of MoE inference is the \emph{sparse and irregular expert activation}, which varies across tokens and layers. 
As a result, most experts remain idle for most of the execution while still occupying memory. 
In edge deployments, keeping all experts in high-bandwidth memory is impractical due to capacity, cost, and power constraints. 
Instead, expert weights are stored in off-package memory, while non-expert weights, activations, and the KV cache remain on-package. 
This creates a hierarchical memory system where expert weights must be fetched over a slower off-package path, such as PCIe. 
During inference, the router selects the top-$k$ experts, triggering on-demand loading of their weights before computation can proceed. 
This introduces a serialization between routing and expert execution, placing expert loading on the critical path.
Due to the size and irregular access of expert parameters, this data movement dominates latency and energy while leaving compute resources idle during the transfer.

Attention computation provides a natural opportunity to overlap expert loading with useful work. If expert usage can be predicted before attention, the corresponding weights can be prefetched while attention is executing, partially or fully hiding expert-loading latency. 
However, imperfect predictions may miss required experts, reintroducing stalls and limiting the effectiveness of naive top-$k$ prefetching, as discussed next.

\subsection{Why Fixed Top-$k$ Prefetch is Insufficient?}\label{ssec:motivation}

While predictive prefetching overlaps expert transfers with attention computation, existing approaches prefetch only the predicted top-$k$ experts.
We show that this strategy is insufficient to reliably eliminate stalls in MoE inference.

\bh{Challenge 1: High overlap accuracy does not imply stall-free execution:}
A key limitation of top-$k$ prefetching is its sensitivity to even small prediction errors. 
If any expert selected by the original router is missing from the prefetched set, it must be fetched on demand, exposing correction latency on the critical path.
Hence, average overlap accuracy across all layers can be misleading. For instance, predicting $k-1$ out of $k$ experts yields an overlap of $\frac{k-1}{k}$ (e.g., 75\% for $k=4$ and 87.5\% for $k=8$). 
Despite these seemingly high values, \emph{tokens at every layer still incur a correction stall}, revealing a fundamental limitation. 
Therefore, we consider per-layer, per-token overlap and complement it with application-level accuracy.

\bh{Challenge 2: Token-level heterogeneity:}
Routing uncertainty varies significantly across tokens and layers. 
Tokens corresponding to common patterns (e.g., frequent words or simple continuations) exhibit predictable routing, while more complex or context-dependent tokens are harder to predict.
A fixed prefetch policy cannot adapt to these variations. 
A small fixed $\delta$ leads to frequent misses on difficult tokens, while a large $\delta$ wastes communication energy on easy ones. 
Therefore, \emph{determining how many additional experts to prefetch is a non-trivial, per-token decision.}

\bh{Challenge 3: Runtime decision must be lightweight:}
Determining the prefetch budget must 
(1) be early enough to maximize overlap with attention and 
(2) introduce minimal area, power, and performance overhead. 
These constraints limit the available decision window and rule out complex or compute-intensive policies.

\begin{figure}[t]
    \centering
    \includegraphics[width=0.9\linewidth]{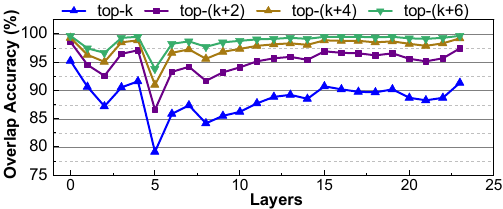}
    \caption{Per-layer overlap accuracy for Granite-3.1-1B-A400M ($k=8$). Increasing the prefetch budget ($\delta$) stabilizes coverage near 100\%, effectively mitigating the routing stochasticity and misprediction stalls.}
    \label{fig:motivation_delta}
\end{figure}

\bh{Key Insight: Small additional prefetch enables near-complete coverage.}
Eliminating stalls requires \emph{full coverage}, i.e., ensuring all experts selected by the original router are present in the prefetched set. 
Our measurements show that most token-layer instances require only a small number of additional experts beyond top-$k$ to achieve this.
Figure~\ref{fig:motivation_delta} illustrates this behavior for the Granite-3.1-1B-A400M model. 
While top-$k$ prefetching results in variable and suboptimal overlap across layers, 
extending the prefetched set to top-$(k+\delta)$ rapidly improves coverage, approaching near-oracle levels with small $\delta$. 
Even modest increases (e.g., $\delta=2$ or $\delta=4$) significantly reduce misses, indicating that \emph{most prediction errors are localized and can be corrected with minimal additional prefetches}.

These observations motivate moving beyond fixed top-$k$ prefetching toward an adaptive strategy that determines the prefetch budget per token and per layer. 
The goal is to select the minimal additional budget needed to achieve high coverage without incurring unnecessary communication overhead. 
APEX achieves this by selecting a token-dependent extra-prefetch budget $\hat{\delta}(x)$ that balances coverage and efficiency.

\begin{figure*}[t]
\centering
    \centering
    \includegraphics[width=0.95\linewidth]{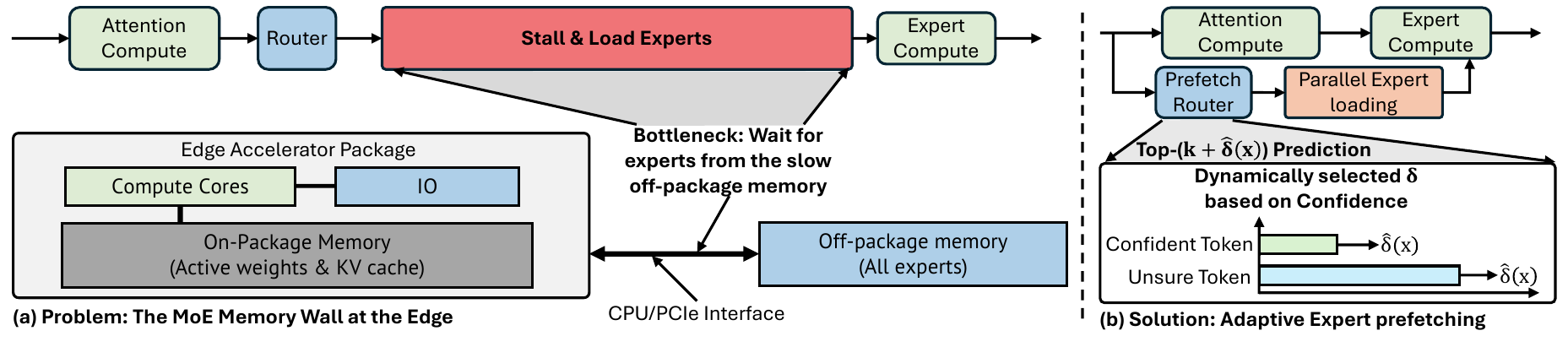}
    \vspace{-3mm}
    \caption{Overview of the MoE memory bottleneck and APEX. (a) Conventional execution stalls while loading experts from off-package memory. (b) APEX overlaps expert loading with attention using a prefetch router that dynamically selects the extra-prefetch budget $\hat{\delta}(x)$ based on token-level confidence.}
    \vspace{-3mm}
    \label{fig:overview}
\end{figure*}

\section{Adaptive Prefetching: Top-$(k+\hat{\delta}(x))$ Expert Prefetching with Logistic CDF}\label{sec:methodology}

\subsection{Overview of Adaptive Expert Prefetching}

Our goal is to reduce exposed expert-loading latency by making expert movement a predictive, token-aware runtime decision.
To achieve this, we augment each MoE layer with a lightweight \emph{prefetch router} placed before the attention block. 
The original router selects the exact top-$k$ experts \textit{after} attention. 
In contrast, the proposed prefetch router runs \textit{before} attention to predict a ranked 
list of candidate experts and initiate data movement while attention is still executing.
This transforms the conventional \texttt{route $\rightarrow$ load $\rightarrow$ execute} pipeline into \texttt{predict $\rightarrow$ prefetch $\rightarrow$ execute}, as shown in Figure~\ref{fig:overview}.

Importantly, APEX does not modify the original routing semantics. 
The prefetch router acts as an auxiliary prediction path that exposes early information for scheduling data movement, while the original router continues to determine the final expert set used for computation. 
This clean separation decouples \emph{expert prediction for memory scheduling} from \emph{expert selection for model execution}.

The key question is how many experts to prefetch as discussed in Section~\ref{ssec:motivation}.
APEX prefetches the top-$(k+\hat{\delta}(x))$ experts, where the additional prefetch budget $\hat{\delta}(x)$ is determined \emph{dynamically} for each token. 
As shown in Figure~\ref{fig:overview}(b), $\hat{\delta}(x)$ depends on token-level confidence.
Easy, high-confidence tokens require only a small prefetch set, while 
uncertain tokens prefetch a few additional experts to improve coverage.

APEX operates in two stages.
\begin{enumerate}[leftmargin=*]
    \item \textbf{Prefetch stage:} The prefetch router produces a ranked expert list, and a confidence model selects $\hat{\delta}(x)$. The runtime then issues asynchronous direct memory access (DMA) transfers for the top-$(k+\hat{\delta}(x))$ experts while attention executes.
        
    \item \textbf{Execution stage:} After attention, the original router determines the exact top-$k$ experts for computation. 
\end{enumerate}

If all routed experts are already loaded to on-package memory, execution proceeds immediately.
Otherwise, APEX uses one of two modes: correctness-preserving mode fetches the missing routed experts while computing the available ones, whereas stall-free mode avoids the correction stall by executing with the available routed experts and high-scoring prefetched substitutes.

\subsection{Problem Formulation and Learning Objective}\label{ssec:objective}

\begin{figure}[t]
    \centering
    \vspace{-3mm}
    \includegraphics[width=0.95\linewidth]{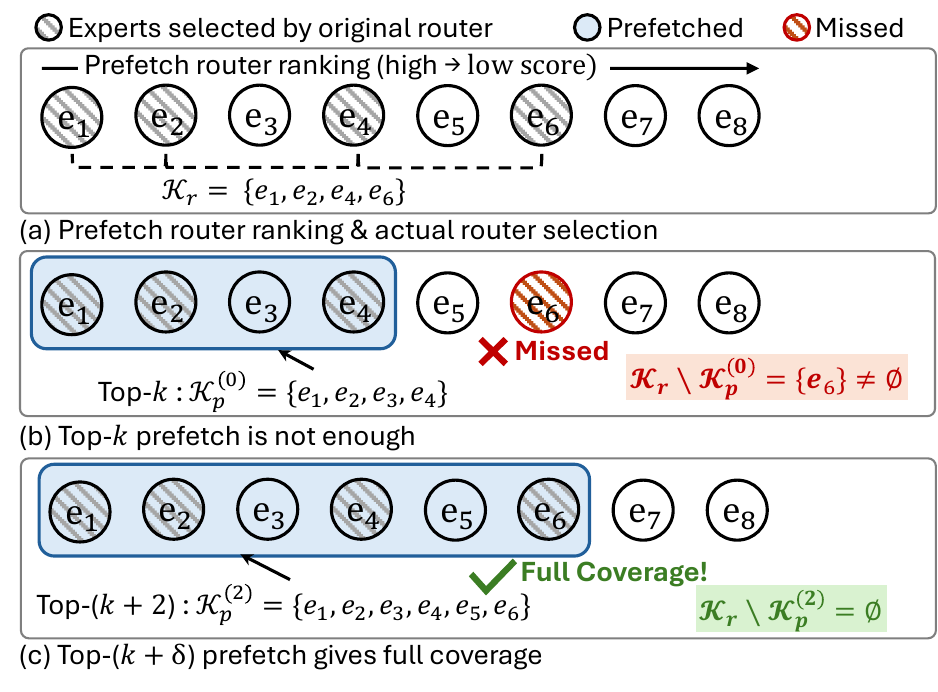}
    \vspace{-3mm}
    \caption{Illustration of expert prefetch coverage. The prefetch router produces a ranked list of $N=8$ experts, while the original router selects $k=4$ experts ($\mathcal{K}_r$). Prefetching only the top-$k$ experts may miss some required experts. Extending the prefetched set to top-$(k+\delta)$ achieves full coverage.}
    \vspace{-3mm}
    \label{fig:problem_formulation}
\end{figure}

Let $\mathcal{E}=\{e_1,e_2,\dots,e_N\}$ denote the ordered list of experts produced by the 
prefetch router, sorted from the highest (most likely to be used) to the lowest predicted score, where $N$ is the total number of experts. 
Let $\mathcal{K}_r \subseteq \mathcal{E}$, with $|\mathcal{K}_r| = k$, denote the set of experts selected by the original router. 
Figure~\ref{fig:problem_formulation}(a) illustrates an example where the prefetch router outputs a ranked list of 8 experts, and the original router selects 4 experts ($k=4$).
For any extra-prefetch budget $\delta \in \{0,1,\dots,N-k\}$, the prefetched expert set is:
\begin{equation}
\mathcal{K}_p^{(\delta)} = \{e_1, e_2, \dots, e_{k+\delta}\}
\end{equation}

A prefetch decision is successful if all required experts are already present in the prefetched set:
\begin{equation}
\mathcal{K}_r \subseteq \mathcal{K}_p^{(\delta)} \quad \Leftrightarrow \quad \mathcal{K}_r \setminus \mathcal{K}_p^{(\delta)} = \varnothing
\end{equation}

As shown in Figure~\ref{fig:problem_formulation}(b), prefetching only the top-$k$ experts (i.e., $\delta=0$) can miss required experts, even when most predictions are correct. 
Extending the prefetched set with a small additional budget $\delta$ can eliminate these misses and achieve full coverage.

We define the \emph{oracle extra-prefetch budget} $\delta^*$ as the minimum $\delta$ that guarantees full coverage:
\begin{equation}
\delta^* = \min \left\{ \delta \in \{0,1,\dots,N-k\} \;\middle|\; \mathcal{K}_r \setminus \mathcal{K}_p^{(\delta)} = \varnothing \right\}
\end{equation}

\begin{figure*}[t]
\centering
    \centering
    \includegraphics[width=0.95\linewidth]{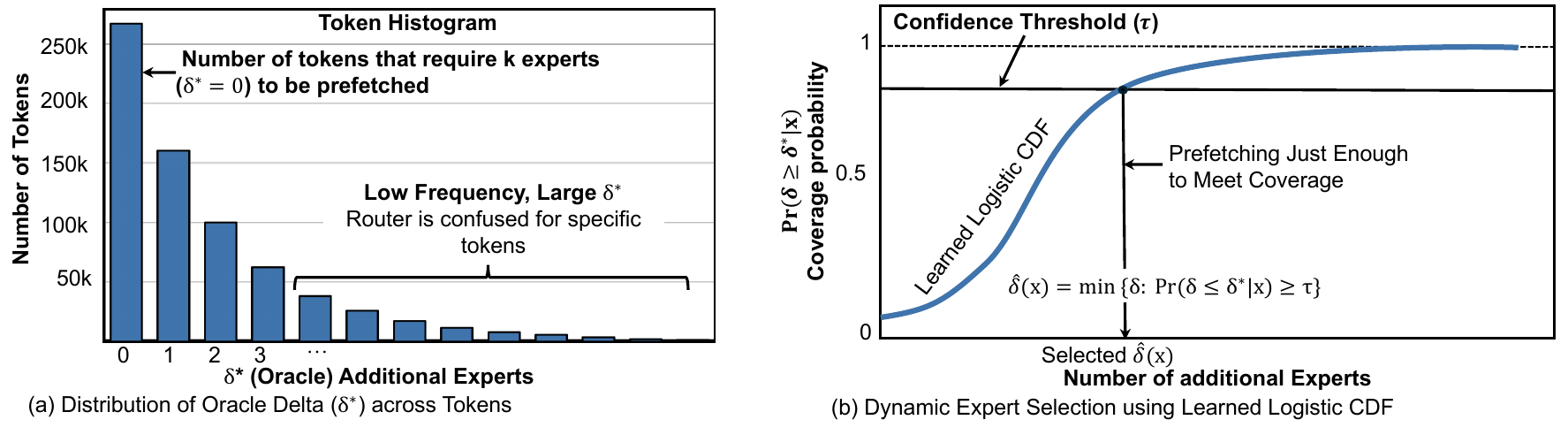}
    \vspace{-2mm}
    \caption{Adaptive expert prefetching. (a) Representative distribution of oracle extra-prefetch budget ($\delta^*$) for Granite-1B (Layer 3) on WikiText~\cite{merity2016wiki}, showing a long-tail across tokens. (b) Selection of prefetch budget $\hat{\delta}(x)$ using a learned logistic CDF to satisfy a target coverage threshold ($\tau$).}
    \label{fig:cdf}
\end{figure*}

Figure~\ref{fig:cdf}(a) shows a representative distribution of $\delta^*$ obtained by profiling the Granite-1B model (Layer 3) on the WikiText~\cite{merity2016wiki} test set. 
Most tokens require little or no additional prefetch, while a small fraction require larger $\delta^*$ due to higher routing uncertainty.

\bh{Learning Objective:}
Our goal is to select a token-dependent prefetch budget $\hat{\delta}(x)$, where x denotes the token representation available before attention, that minimizes unnecessary data movement while ensuring high coverage probability.
Formally, we require that the prefetched set covers the original router's selection with probability at least $\tau$:
\begin{equation}
\Pr\left(\mathcal{K}_r \setminus \mathcal{K}_p^{(\hat{\delta}(x))} = \varnothing \mid x\right) \ge \tau,
\end{equation}
where $\tau \in [0,1]$ is a user-defined coverage target.

Since full coverage is achieved if and only if $\hat{\delta}(x)  \ge \delta^*$, the objective can be equivalently written as:
\begin{equation}
\Pr\left(\hat{\delta}(x) \ge \delta^* \mid x\right) \ge \tau
\end{equation}

This formulation highlights the core objective: for each token, select the smallest token-dependent $\hat{\delta}(x)$ that satisfies the coverage constraint, thereby balancing latency reduction against communication cost.

\vspace{-2mm}
\subsection{Confidence Modeling via CDF}\label{ssec:cdf}

The adaptive prefetching problem reduces to selecting the smallest extra-prefetch budget that satisfies a target coverage probability. 
To enable this decision at runtime, we use a lightweight probabilistic model that estimates whether a given number of additional experts is sufficient.

\bh{Coverage Probability:}
We define the probability that a candidate budget $\delta$ is sufficient as:
\begin{equation}
p_\delta(x) = \Pr(\delta \ge \delta^* \mid x), \quad \delta \in \{0,\dots,N-k\}
\end{equation}

Since larger prefetch budgets can only improve coverage, these probabilities are monotonic:
\begin{equation}
p_0(x) \le p_1(x) \le \cdots \le p_{N-k}(x),
\end{equation}
which can be interpreted as the cumulative distribution function (CDF) of the oracle $\delta^*$.
Figure~\ref{fig:cdf}(b) provides intuition for this formulation. 
The learned CDF enables APEX to select the smallest $\delta$ that satisfies a target threshold $\tau$.

\bh{Ordinal Logistic CDF Model:}
We model this CDF using an ordinal logistic formulation that captures the ordered structure of $\delta$. 
The cumulative probability is modeled by:
\begin{equation}
p_\delta(x) = \sigma(\theta_\delta - w^\top x), \quad \delta \in \{0,\dots,N-k\},
\end{equation}
where $w$ is a learned parameter vector, $\{\theta_\delta\}$ are ordered bias terms satisfying $\theta_0 \le \theta_1 \le \cdots \le \theta_{N-k}$, and $\sigma(\cdot)$ is the sigmoid function.

The ordered biases partition the latent confidence space, with each threshold corresponding to a different prefetch budget.
By construction, this formulation guarantees monotonicity of $p_\delta(x)$ and therefore defines a valid CDF.

\begin{table}[t]
\centering
\vspace{-1mm}
\caption{Summary of notations used in this work.}
\vspace{-1mm}
\label{tab:notation}

\resizebox{\linewidth}{!}{%
\begin{tabular}{cl}
\toprule
\textbf{Notation} & \textbf{Description} \\
\midrule
$\mathcal{E} = \{e_1, e_2, \dots, e_N\}$  & Ordered list of experts sorted by prefetch router\\
$N$ & Total number of experts \\
$k$ & Number of experts to be executed \\
$\delta \in \{0,1,\dots,N-k\}$ & Extra-prefetch budget \\
$\mathcal{K}_r$ & Expert set selected by the original router \\
$\mathcal{K}_p^{(\delta)}$ & Prefetched expert set $\{e_1, \dots, e_{k+\delta}\}$ \\
$\delta^*$ & Oracle delta: minimum $\delta$ such that $\mathcal{K}_r \subseteq \mathcal{K}_p^{(\delta)}$ \\
$\hat{\delta}(x)$ & Token-dependent (dynamic) prefetch budget \\
$\tau \in [0,1]$ & Target coverage probability threshold \\
$p_\delta(x)$ & Coverage probability: $p_\delta(x) = \Pr(\delta \ge \delta^* \mid x)$ \\
$w$ & Learned parameter vector in CDF model \\
$\theta_0 \le \theta_1 \le \cdots \le \theta_{N-k}$ & Ordered thresholds (cutpoints) for ordinal model \\
\bottomrule
\vspace{-3mm}
\end{tabular}
}
\end{table}

At runtime, given a token representation $x$ and a target coverage $\tau$, we select the smallest prefetch budget that satisfies:
\begin{equation}
\hat{\delta}(x) = \min \left\{ \delta \in \{0,\dots,N-k\} \;\middle|\; p_\delta(x) \ge \tau \right\}.
\end{equation}
This yields a confidence-aware adaptive policy: 
\begin{itemize}[leftmargin=*]
    \item easy tokens (tokens whose routed experts are likely covered by a small budget) require a small $\hat{\delta}(x)$;
    \item difficult tokens with low confidence require a larger $\hat{\delta}(x)$.
\end{itemize}

Thus, unlike static overfetching, APEX fetches \emph{just enough} additional experts to satisfy the desired coverage target, reducing misses while avoiding unnecessary communication.

\vspace{-1mm}
\subsection{Prefetch Router: Training and Integration}\label{ssec:training}
\vspace{-0.5mm}
The key design goal is to enable accurate early prediction for prefetching \emph{without modifying or retraining the base MoE model}. 
Accordingly, the LLM remains frozen, and training is limited to the auxiliary prefetching components, i.e., the prefetch router and the CDF model.

\bh{Prefetch Router Architecture and Placement:}
The prefetch router mirrors the original MoE router: a linear layer that produces expert logits, followed by a softmax to produce a probability distribution over experts. 
It is placed before the attention block of each MoE layer and operates on the layer’s hidden representation.
This placement serves two purposes. 
First, it hides expert loading latency by initiating expert transfers during attention. 
Second, it remains within the same layer, preserving a strong correlation between the prefetch prediction and the final routing decision made later by the original router.

\bh{Prefetch Router Distillation:}
For each MoE layer, let $q_r$ and $q_p$ denote the expert softmax distributions produced by the original router and the prefetch router, respectively. 

The prefetch router is trained to match the original router using a KL divergence loss:
\begin{equation}
\mathcal{L}_{\mathrm{KL}} = \sum_{i} q_r(i)\log \frac{q_r(i)}{q_p(i)}.
\end{equation}

Training requires only forward passes through the frozen base model to obtain $q_r$, and gradients are applied exclusively to the prefetch router. 
\textit{No changes are made to expert execution during this stage}. 
The result is a layer-wise early predictor that approximates the original router's expert selections.

\bh{CDF Model Training:}\label{bh:cdf}
After training the prefetch router, we fit the CDF model described in Section~\ref{ssec:cdf} through a separate lightweight stage. 
For each token and layer, we use the trained prefetch router to obtain expert probabilities, derive the ranked expert list, and compute the oracle budget $\delta^*$ by comparing with the original router. 
These are converted into cumulative binary targets $\mathbb{I}[\delta \ge \delta^*]$ for each $\delta$.

The CDF model is trained using a cumulative binary cross-entropy loss:
\begin{equation}
\mathcal{L}_{\text{CDF}} = \sum_{\delta=0}^{N-k} \mathcal{H}\big(p_\delta(x),~ \mathbb{I}[\delta \ge \delta^*]\big),
\end{equation}
where $\mathcal{H}(\cdot,\cdot)$ denotes binary cross-entropy~\cite{goodfellow2016deep}.

The model consists of a single weight vector and a small set of thresholds, making training inexpensive and feasible with a modest held-out validation set.
Since the prefetch router is distilled from the frozen original router, it is tied to the base model’s routing function rather than to a specific prompt or downstream task. 
Thus, the same trained prefetch router and CDF model can be reused across prompts and datasets; retraining is only needed if the base MoE model is fine-tuned and its router behavior changes.

\bh{Runtime Operation:}
The prefetch router and CDF model are evaluated before attention to determine the adaptive prefetch budget $\hat{\delta}(x)$. 
The system then asynchronously prefetches the top-$(k+\hat{\delta}(x))$ experts while attention executes. 
After attention, the original router produces the final top-$k$ selection, and execution proceeds using either correction or stall-free mode, as described next.
The overhead is minimal, as detailed in Section~\ref{ssec:overhead}.  
The prefetch router mirrors the lightweight original router, and the CDF model introduces only a small number of parameters, enabling early prediction and scheduling with negligible compute and storage overhead.

\subsection{Correctness-preserving and Stall-free Execution Modes}\label{ssec:varients}

Even with adaptive top-$(k+\hat{\delta}(x))$ prefetching, rare mispredictions may occur. 
Hence, we support two execution modes. 

\bh{1) Correctness-preserving mode:} 
The system strictly follows the original router decision.
If any experts in $\mathcal{K}_r$ are missing from the prefetched set, they are fetched while execution begins on available experts, and the final output uses the complete original expert set.
This preserves exact MoE semantics, but incurs a small residual correction latency, which is evaluated in Section~\ref{ssec:perf}.

\bh{Asynchronous miss correction:}\label{bh:miss_correction}
APEX does not use a rigid wait-then-execute pipeline when a prefetch miss occurs. 
Since the selected experts in an MoE layer are independent before the final weighted sum, correctness-preserving execution begins immediately on the routed experts that are already available in the on-package buffer. 
Missing routed experts are fetched asynchronously in parallel with this available expert computation. 
After the missing experts arrive, APEX computes only the remaining expert outputs and then performs the final weighted aggregation using the original router weights. 
Therefore, the correction overhead is the unhidden portion of the miss transfer after overlap with available expert computation, rather than the full load latency. 
This preserves exact MoE semantics while exploiting inter-expert pipelining.

\bh{2) Stall-free mode:}
This mode eliminates correction latency by executing only with experts that are already present in the prefetched set.
Instead of waiting to fetch missing experts, we approximate the original routing operation using the available experts.
If all experts selected by the original router are already prefetched (Algorithm~\ref{alg:nocorrect}, lines 3--4), we directly execute them without any deviation from the behavior of the original architecture.
Otherwise, if one or more routed experts are missing, stall-free mode keeps the correctly prefetched routed experts and replaces the missing experts with the same number of highest-weight candidates from the remaining prefetched set, ranked by the original router's softmax weights (lines 7--11).

Intuitively, this procedure selects the highest-scoring available substitutes from the prefetched set according to the original routing scores, minimizing deviation from the original decision. Because overlap is already very high, such substitutions are rare. 
Moreover, the missing expert are often a lower-ranked choice, so the resulting approximation can have small model-dependent accuracy impact, as evaluated in Section~\ref{sssec:application}. 
We note that, the stall-free mode is an optional mode, while correctness-preserving mode remains the default when exact routing semantics are required.

\begin{algorithm}[t]
\caption{Expert Selection in the Stall-free Mode}
\label{alg:nocorrect}
\begin{algorithmic}[1]
\STATE \textbf{Input:} Routed experts $\mathcal{K}_r$, original router logits $q_r \in \mathbb{R}^{N}$, prefetched experts $\mathcal{K}_p^{(\hat{\delta})}$
\STATE \textbf{Output:} Execution experts $\tilde{\mathcal{K}}$

\STATE $\alpha \leftarrow \mathrm{softmax}(q_r)$ \textit{// original-router expert weights}
\IF{$\mathcal{K}_r \subseteq \mathcal{K}^{(\hat{\delta})}_p$}
\STATE $\tilde{\mathcal{K}} \leftarrow \mathcal{K}_r$ \textit{// all correct experts are available}
\ELSE
\STATE $\tilde{\mathcal{K}} \leftarrow \mathcal{K}_r \cap \mathcal{K}^{(\hat{\delta})}_p$ \textit{// keep correctly routed experts}
\STATE $\mathcal{M} \leftarrow \mathcal{K}_r \setminus \mathcal{K}^{(\hat{\delta})}_p$ \textit{// missing routed experts}
\STATE $\mathcal{C} \leftarrow \mathcal{K}^{(\hat{\delta})}\setminus \tilde{\mathcal{K}}$ \textit{// available replacement candidates}
\STATE $\mathcal{S} \leftarrow \mathrm{Top}*{|\mathcal{M}|}(\mathcal{C}; \alpha)$ \textit{// highest-weight candidates}
\STATE $\tilde{\mathcal{K}} \leftarrow \tilde{\mathcal{K}} \cup \mathcal{S}$
\ENDIF
\STATE \textbf{return} $\tilde{\mathcal{K}}$
\end{algorithmic}
\end{algorithm}

\section{Experimental Evaluation}\label{sec:eval}

\subsection{Experimental Setup}\label{ssec:setup}
\bh{Models:}
We evaluate APEX on four representative MoE LLMs spanning different model scales and configurations:
\begin{itemize}[leftmargin=*]
    \item \textit{Granite-1B} (IBM Granite-3.1-1B-A400M~\cite{granite31_1b_a400m}): a 1B-scale model with $N=32$ experts and top-$k$ routing with $k=8$.
    \item \textit{Granite-3B} (IBM Granite-3.1-3B-A800M~\cite{granite31_a800m}): a 3B-scale model with $N=40$ experts and top-$k$ routing with $k=8$.
    \item \textit{Phi-7B} (Microsoft Phi-mini-MoE-7B-A2.4B~\cite{phi3_technical_report}): a 7B model with $N=16$ experts and top-$k$ routing with $k=2$.
    \item  DeepSeek-16B (DeepSeek-V2-Lite-16B-A2.4B)~\cite{liu2024deepseekv2}: a 16B model with $N=64$ experts and top-$k$ routing with $k=6$.
\end{itemize}

These models capture varying numbers of experts, routing sparsity, and layer depths. 
Granite-1B, Granite-3B, Phi-7B, and DeepSeek-16B contain 24, 32, 32, and 27 Transformer layers, respectively. 
In DeepSeek-16B, the first Transformer layer uses a dense FFN, while the remaining 26 layers use MoE FFNs. 
DeepSeek also includes two shared experts per MoE layer that are activated for every token in addition to the six routed experts selected from the 64 routed experts. 
Thus, for DeepSeek-16B, shared experts are deterministic per-token computation and are not part of APEX's adaptive routing or extra-prefetch-budget decision.

\bh{Workload:}
We focus on the decode phase (autoregressive token generation), since it is latency-critical in edge deployments and exposes expert-loading bottlenecks. 
We evaluate across diverse context lengths, including 512, 1024, and 2048 tokens. 
For application-level accuracy, we use standard benchmarks including WikiText~\cite{merity2016wiki} (perplexity), AI2 Reasoning Challenge~\cite{clark2018think}, Massive Multitask Language Understanding~\cite{hendrycks2020measuring}, WinoGrande~\cite{sakaguchi2021winogrande}, and TruthfulQA~\cite{lin2022truthfulqa}.

\bh{Prefetch Router and CDF Training:}\label{bh:cdf_training}
Training is performed on an NVIDIA RTX 3090 GPU~\cite{NVIDIA_RTX3090} for Granite and Phi models and A100~\cite{NVIDIA_A100} for DeepSeek using the WikiText dataset~\cite{merity2016wiki}. 
We use a learning rate of $5\times10^{-4}$, batch size of 8, sequence length of 1024, and train for 1000 steps. 
The training takes approximately 10 minutes for Granite-1B, 30 minutes for Granite-3B, and about 1 hour for Phi-7B and DeepSeek-16B. 
After distillation, the CDF model is fitted using collected oracle $\delta^*$ values on a held-out validation set.
No task-specific retraining or hyperparameter search is used for downstream benchmarks; the auxiliary components trained on WikiText are reused for all evaluated application-level tasks.

\bh{Baselines:}
We compare APEX against three baselines:
(i) \emph{No Prefetch}, the conventional execution scheme in which expert loading occurs on the critical path~\cite{eliseev2023fast};
(ii) \emph{ProMoE}, a state-of-the-art prediction-based top-$k$ prefetching approach~\cite{song2024promoe};
and (iii) \emph{static overfetch} policies with fixed additional-prefetch budgets. 
We evaluate both \emph{correctness-preserving} and \emph{stall-free} modes.
To decouple the effect of expert prediction from adaptive budget selection, the static overfetch baselines use the same prefetch-router ranking as APEX and replace the learned ($\hat{\delta}(x)$) with a fixed ($\delta$) for every token and layer. 
Therefore, differences between static overfetching and APEX isolate the benefit of adaptive per-token budgeting.

The evaluation metrics are (i) \emph{overlap accuracy} between prefetched and routed experts, (ii) \emph{application-level accuracy} across benchmarks, and (iii) \emph{system efficiency}, in terms of latency, energy, and energy-delay-product (EDP). 

\subsection{Hardware Platform and Evaluation Methodology}\label{ssec:hw}

\begin{figure}[t]
    \centering
    \includegraphics[width=0.8\linewidth]{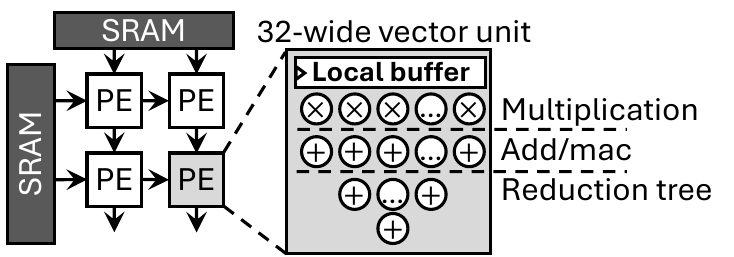}
    \caption{Microarchitecture of a vector processing array.}
    \label{fig:compute_array}
\end{figure}

\bh{Target Edge Architecture:}
We model a modern edge-class accelerator consisting of a compute chiplet tightly coupled with I/O and memory chiplets, following the prior approach~\cite{shan2022architecture}. 
The compute chiplet integrates multiple vector processing arrays and SRAM buffers~\cite{alish2026duet}, while all expert weights reside in off-package LPDDR5X memory.
The compute microarchitecture is based on the vector processing array illustrated in Figure~\ref{fig:compute_array}. 
Each array is organized as a $16\times16$ grid of vector units with vector length 32, supporting bfloat16 (BF16) matrix-multiply operations. 
We instantiate four such arrays operating at 750~MHz, achieving a peak throughput of 24 TFLOPS, as summarized in Table~\ref{tab:hw_config}. 
Each array is provisioned with 2 MB of SRAM (8 MB total) and includes 32 special-function units~\cite{reggiani2023flex} to support non-linear activations.
We use BF16 as the default evaluation datatype to preserve the original model behavior and isolate the impact of prefetching; Section~\ref{ssec:ablation} further studies how APEX behaves when expert weights are represented with lower-precision datatypes.

\begin{figure*}[t]
\centering
    \includegraphics[width=\linewidth]{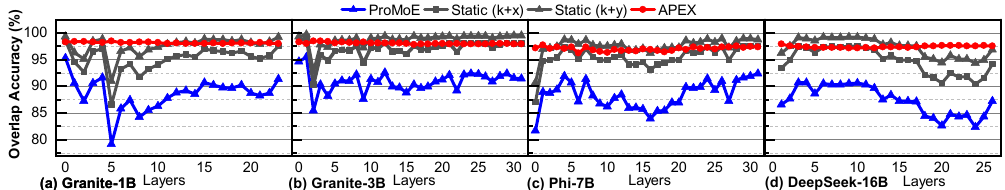}
    \caption{Per-layer overlap accuracy using a CDF confidence threshold of $\tau=0.90$ for APEX.
    \textbf{(a)} Granite-1B model. 
    The static baselines prefetch top-($k+2$) and top-($k+4$), while APEX prefetches on average 4.17.
    \textbf{(b)} Granite-3B model (keeping the same static-baseline settings). 
    \textbf{(c)} Phi-7B model. The static baselines prefetch top-($k+1$) and top-($k+2$), while APEX prefetches on average 0.67.
    \textbf{(d)} DeepSeek-16B model. The static baselines prefetch top-($k+2$) and top-($k+4$), while APEX prefetches on average 1.98.}
    \label{fig:layer_overlap}
\vspace{-3mm}
\end{figure*}

To reflect realistic edge constraints, on-package memory capacity is varied across model scales (1~GB, 2~GB, and 6~GB), holding only non-expert weights, activations, and KV cache, following the literature~\cite{kwon2023efficient,aminabadi2022deepspeed}. 
All expert parameters are stored in off-package LPDDR and accessed over a PCIe interconnect. 
On-package memory bandwidth is modeled at 819~GB/s (HBM3-class), while off-package bidirectional bandwidth is set to 256~GB/s (PCIe 6.0 x16)~\cite{pcie6_spec,park2022192}.
The evaluations are performed on a latency-sensitive single-user edge inference setting, but the proposed method supports batching multiple concurrent requests.

\bh{Platform Positioning:}\label{bh:platform}
The target configuration represents a high-end, near-future edge AI accelerator rather than a low-power smartphone-class SoC. 
This class is motivated by emerging robotics, industrial, and local GenAI platforms that already integrate high compute throughput with large local memory capacity, such as NVIDIA Jetson AGX Orin/Thor~\cite{nvidia_jetson_thor}, and Hailo-10H~\cite{hailo_10h_accelerator} class edge accelerators. 
At the same time, APEX is not tied to this specific bandwidth point. 
Section~\ref{ssec:ablation} evaluates sensitivity to lower and higher off-chip bandwidths, including a 32 GB/s setting representative of more constrained mobile-SoC-class memory systems.

\bh{Evaluation Methodology:}\label{bh:eval}
Our evaluation combines hardware synthesis with cycle-accurate co-simulation rather than pure simulation. 
The compute arrays are implemented in RTL, synthesized in TSMC 28 nm, and their active/leakage power is characterized using Synopsys PrimeTime~\cite{primetime}. 
Memory timing is modeled using Ramulator, which captures DRAM bank conflicts, refresh, and controller scheduling~\cite{luo2023ramulator}, while LPDDR and PCIe energy costs are incorporated using established per-bit energy models~\cite{ghose2018dram,pcie6_spec}.
The arrays are clock-gated, allowing us to accurately model idle periods during stalls. 
We obtain \emph{active power} by running the arrays under high activity, and \emph{leakage power} by evaluating the design under no-activity conditions with clock gating enabled. 
On-package communication is modeled using UCIe-based interconnect parameters that capture low-energy, high-bandwidth chiplet communication~\cite{sharma2022universal}.
We use an open-source, cycle-accurate compute–communication co-simulation framework, CHIPSIM~\cite{pfromm2025chipsim}, which integrates:
(i) compute timing from RTL-derived models,
(ii) memory access timing from DRAM simulation~\cite{luo2023ramulator},
and (iii) per-layer execution traces collected from GPU-based model runs~\cite{NVIDIA_RTX3090}. 
The simulator models token-level execution, including attention compute, expert routing, data transfers, and expert execution, enabling us to model overlap among attention computation, expert routing, DMA transfers, and expert execution.
The asynchronous expert transfers are not modeled as ideal parallel fetches. 
Each prefetched expert is injected as a DMA transfer into the shared PCIe/LPDDR path, where concurrent requests experience bandwidth limits and queuing delay. 
CHIPSIM coordinates compute, routing, DMA transfers, and expert execution on a unified timeline. Thus, overlap is limited by the modeled attention window and contention in the off-chip memory path.

\begin{table}[t]
\centering
\caption{Hardware configuration and power parameters}
\label{tab:hw_config}
\resizebox{\linewidth}{!}{%
\begin{tabular}{lcc}
\toprule
\textbf{Component} & \textbf{Configuration} & \textbf{Perf./Energy} \\
\midrule
Compute Arrays & $4 \times (16\times16\times32)$ & 24 TFLOPS @ 750 MHz \\
SRAM~\cite{balasubramonian2017cacti} & 8 MB (2 MB per array) & CACTI-based modeling \\
HBM Bandwidth~\cite{park2022192} & 819 GB/s & 7 pJ/bit \\
LPDDR5X ~\cite{ghose2018dram} & Off-package experts & 3 pJ/bit \\
PCIe 6.0 x16 ~\cite{pcie6_spec} & 256 GB/s & 5 pJ/bit \\
On-chip Comm.~\cite{sharma2022universal} & UCIe Adv. & 0.5 pJ/bit \\
\bottomrule
\end{tabular}
}
\vspace{-2mm}
\end{table}

\bh{Prefetch-overlap Bound:}~\label{bh:bound}
APEX does not assume that all prefetched experts are fully hidden behind attention. For an extra-prefetch budget ($\delta$), the ideal hiding condition is:
\begin{equation}
T_{\mathrm{prefetch}}(\delta)=\frac{\delta S_{\mathrm{expert}}}{B_{\mathrm{eff}}}+T_{\mathrm{queue}} \leq T_{\mathrm{attn}},
\end{equation}
where ($S_{\mathrm{expert}}$) is the expert size, ($B_{\mathrm{eff}}$) is the effective off-chip bandwidth, and ($T_{\mathrm{queue}}$) captures memory-controller/interconnect contention. When this condition is not satisfied, the unhidden portion of ($T_{\mathrm{prefetch}}$) appears as exposed stall time in the simulation. 
Thus, APEX improves latency and energy only when the saved stall/idle cost exceeds the additional transfer cost. This trade-off is evaluated directly through CHIPSIM and further stressed in the bandwidth sensitivity study in Section~\ref{ssec:ablation}.

\bh{Methodology Scope and Limitations:}\label{bh:scope}
CHIPSIM has been validated against hardware measurements on a real chiplet platform under concurrent workloads. However, the target APEX accelerator itself is evaluated through synthesis-calibrated co-simulation rather than measurements on a hardware prototype.
We model PCIe/LPDDR contention through the memory/interconnect timing model and account for OS-level DMA setup and Input-Output Memory Management Unit (IOMMU) overhead using a fixed per-DMA-descriptor cost, which is amortized over multi-MB expert transfers.
Building an FPGA or silicon prototype to further validate these system-level effects and full end-to-end OS-stack modeling are left as future work.

\subsection{Expert Prediction Overlap Accuracy}\label{ssec:accuracy}
This section analyzes the \textit{overlap accuracy} between the experts selected by the original router and the prefetched top-$(k+\hat{\delta}(x))$ set, since a high overlap is the key enabler for hiding expert-loading latency. 
Then, the next subsection analyzes the impact of the rare routing mismatches in the stall-free mode on the end-to-end \emph{application-level accuracy}.

\bh{Per-layer Overlap Across Models:}
We first evaluate how the overlap accuracy changes across layers and model scales to ensure that APEX consistently offers high coverage throughout the network. 
In this experiment, the confidence threshold is set as $\tau=0.90$ to ensure a high and consistent overlap target across all layers. 
The effect of varying this threshold and its impact on overlap accuracy is analyzed later.

Figure~\ref{fig:layer_overlap} plots the per-layer overlaps between the original router's selections and the prefetched experts for Granite-1B, Granite-3B, Phi-7B, and DeepSeek-16B.
Across all models, APEX consistently maintains over $97\%$ overlap across layers, greatly reducing the variability seen in alternative methods. 
ProMoE exhibits noticeable degradation in layers 5 and 8 of Granite-1B, with $\sim$79\% and $\sim$84\% overlap, respectively.
Similarly, it struggles with multiple layers in the Granite-3B, Phi-7B and DeepSeek-16B models, 
where ProMoE often stays in the high-80\% range and drops near 82\%. 
In contrast, APEX consistently achieves around 97–98\% overlap.
This stability is essential to guarantee that no individual layer acts as a bottleneck, effectively removing layer-wise performance differences caused by mispredictions.

Figure~\ref{fig:layer_overlap} also shows that APEX significantly outperforms static policies that always prefetch 2 and 4 extra experts. 
Fixed policies must account for worst-case layers, which causes unnecessary overfetching in simpler layers. Even with this over-provisioning, they struggle to prefetch the correct set of experts for many layers (e.g., 5$^\text{th}$ layer of Granite-1B and first layer of Phi-7B), which likely exhibit higher routing ambiguity and weaker correlation between the pre-attention representation and the final routing decision. 
In contrast to static policies, APEX adapts its prefetch budget to runtime uncertainty, eliminating the low-overlap tail while avoiding excessive overfetch. 
As a result, APEX maintains over $97\%$ overlap across layers by prefetching  4.17 (Granite-1B), 2.86 (Granite-3B), 0.67 (Phi-7B) 1.98 (DeepSeek-16B) experts on average.
These results show that APEX enhances not just average overlap but also worst-case reliability, which is essential for latency-sensitive inference.

\begin{figure}[b!]
\centering
    \centering
    \includegraphics[width=\linewidth]{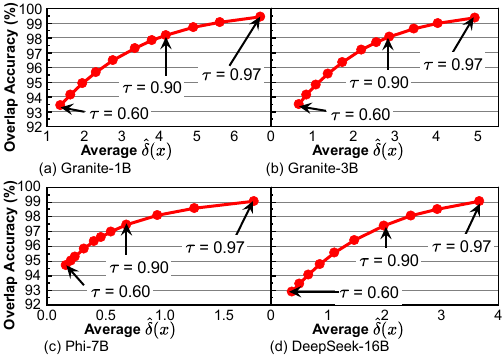}
    \caption{Average overlap accuracy vs. extra-prefetch budget ($\hat{\delta}$). Increasing the CDF confidence threshold ($\tau$) from 0.60 to 0.97 consistently raises overlap toward 100\% across all models, validating that the CDF-based selection effectively meets the desired coverage targets.}
    \label{fig:cdf_delta}
\end{figure} 

\bh{Overlap vs. Adaptive Target Threshold:}
Next, we evaluate how the target confidence threshold $\tau$ affects the overlap accuracy and prefetch budget. 
Figure~\ref{fig:cdf_delta} shows that increasing $\tau$ monotonically increases the overlap percentage across all models, as expected. 
For Granite-1B, the overlap improves from 93.4\% with $\tau=0.60$ to 98.2\% with $\tau=0.90$, and further to 99.4\% with $\tau=0.97$, while the average $\hat{\delta}(x)$ increases from 1.36 to 4.17 and 6.70, respectively. 
Granite-3B shows a similar trend, with overlap increasing from 93.5\% to 98.1\% and 99.4\%, while the average number of prefetched experts increases from $\hat{\delta}(x)$ of 0.67 to 2.86 and 4.93. 
Similarly, for Phi-7B, the average overlap percentage increases from 94.7\% to 97.5\%, and then 99.0\%, while prefetching only 0.16, 0.67, and 1.75 additional experts on average.
Finally, DeepSeek-16B also follows the same pattern, with overlap increasing from 92.9\% to 97.4\% and 99.0\%, while the average prefetched experts increases from 0.37 to 1.98 and 3.66.

These results highlight two key insights. 
First, the CDF-based selector is well-designed: higher $\tau$ consistently yields higher overlap, confirming that APEX can reliably meet the coverage targets. 
Second, near-complete overlap ($>$99\%) can be achieved with a modest increase in prefetch budget, indicating that APEX efficiently allocates additional experts only when necessary rather than uniformly overfetching.
We use $\tau=0.90$ in the remaining experiments because it captures most of the overlap gains before saturation. Further increase causes additional prefetching with only marginal benefits.

\begin{figure}[t]
    \centering
    \includegraphics[width=1\linewidth]{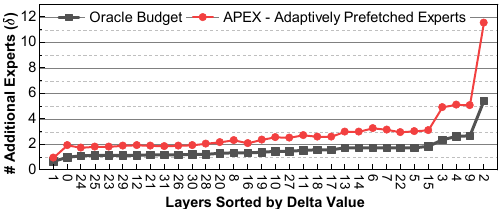}
    \caption{Per-layer additional experts selected by APEX ($\hat\delta(x)$) versus oracle extra-prefetch budget ($\delta^*$) for Granite-3B ($\tau=0.90$). Layers are ordered by oracle $\delta^*$ to highlight relative difficulty.}
    \vspace{-3mm}
    \label{fig:oracle_vs_apex}
\end{figure}

\bh{Adaptive Behavior Compared to Oracle:}
Next, we analyze how effectively APEX adapts its prefetch budget to the intrinsic difficulty of each layer by comparing it to an Oracle that stores the ground truth.
Figure~\ref{fig:oracle_vs_apex} plots the number of required experts (given by the Oracle) and prefetched by APEX with $\tau=0.90$ for the representative Granite-3B model. 
We observe that APEX closely follows the ground truth. 
APEX prefetches three or fewer experts for most layers, which require one or two additional experts according to the Oracle. 
That is, it learns the layers for which experts can be predicted accurately and assigns small budgets to them. 
In more challenging layers that require three or more additional experts, APEX slightly overestimates the oracle budget, which is desirable to provide a safety margin to maintain high overlap.

We observe the same adaptive behavior across other benchmark models. 
Overall, these results confirm that APEX does not use a coarse global heuristic but dynamically adjusts its prefetch budget in response to layer- and token-level runtime variability, enabling it to achieve consistently high overlap accuracy while remaining efficient.
Hence, APEX achieves its primary goal of \emph{transforming a highly variable expert prediction problem into a controlled, confidence-driven prefetch policy}, which consistently delivers very high overlap accuracy while avoiding the inefficiencies of static overfetching.

\begin{figure}[t]
    \centering
    \includegraphics[width=1\linewidth]{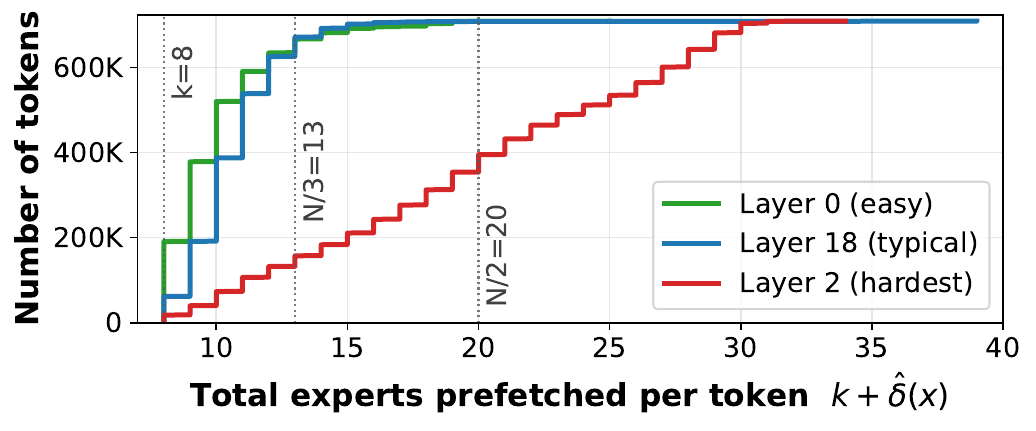}
    \caption{Distribution of total prefetched experts per token ($k+\hat{\delta}(x)$) for Granite-3B at $\tau=0.90$.}
    \label{fig:prefetch_budge}
\end{figure}

\bh{Per-token Prefetch Budget Distribution:}\label{bh:token_dist}
Beyond average ($\hat{\delta}(x)$), we also examine the per-token distribution of the total prefetched experts ($k+\hat{\delta}(x)$), since large bursts could stress the PCIe path or on-package buffers. 
Figure~\ref{fig:prefetch_budge} plots the token distribution for Granite-3B at ($\tau=0.90$) across representative easy, typical, and hardest layers. 
For easy and typical layers, the distribution is concentrated near (k), with mean ($\hat{\delta}$) below 3 and fewer than 0.2\% of tokens prefetching more than (N/2) experts. 
The hardest layer requires a larger budget because the pre-attention representation is less correlated with the final router decision. 
However, APEX prefetches experts layer by layer; each layer's experts are loaded, used, and evicted before the next layer begins, so peak buffer occupancy is bounded by the maximum single-layer prefetch set rather than accumulating across layers. 
Thus, the adaptive budget does not create an unbounded memory-pressure.

\begin{figure*}[b]
    \centering
    \includegraphics[width=1\linewidth]{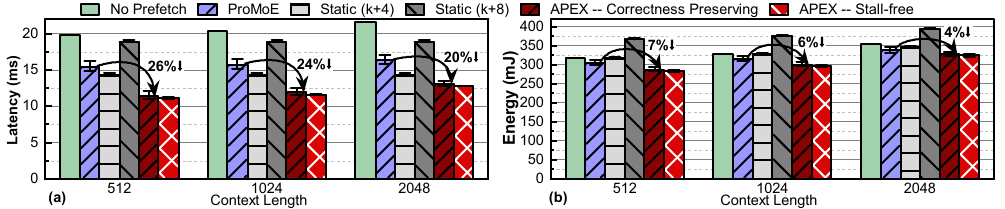}
    \caption{Comparison of (a) average per-token latency and (b) per-token energy consumption across context lengths for Granite-3B model.} 
    \label{fig:perf_energy_latency}
\end{figure*}

\subsection{Application-Level Accuracy}\label{sssec:application}
This section evaluates the impact of the APEX on the application-level accuracy measured by perplexity and standardized benchmarks AI2 Reasoning Challenge (ARC), Massive Multitask Language Understanding (MMLU), WinoGrande (WG), and TruthfulQA (TQA). 
For clarity, we report the perplexity and the average of these standardized scores.
These accuracy metrics are compared against the base system, which always uses the original router's expert choices. 

The correctness-preserving mode always matches the accuracy of the baseline by definition. 
Therefore, Table~\ref{tab:application_acc} reports perplexity and downstream task performance across four model scales under different confidence thresholds $\tau$.
The impact on the perplexity is negligible for the Granite-1B, 3B and DeepSeek-16B models, especially at higher thresholds. 
Specifically, for Granite-1B moving from correctness-preserving execution to stall-free at $\tau=0.90$ results in only a marginal increase in perplexity (7.88 $\rightarrow$ 7.99) and a small drop in average downstream accuracy (43.3 $\rightarrow$ 42.8). 
A similar trend is observed in Granite-3B, where the average score decreases modestly from 54.0 to 53.2, with perplexity increasing slightly from 6.79 to 6.83.
DeepSeek-16B follows the same trend, with only a small PPL increase at $\tau=0.90$ (7.02 $\rightarrow$ 7.10) and a modest average-score drop ( 51.6 $\rightarrow$ 50.8).
These results indicate that in high-overlap regimes, the effect of occasional expert mispredictions is negligible for Granite models.

The robustness in accuracy can be attributed to the routing structure of Granite and DeepSeek models, where each token is processed by multiple experts ($k=8$ for Granite and $k=6$ plus two shared experts for DeepSeek) and the final output is a weighted average.
In this setting, mispredictions are more likely to occur among lower-ranked experts that contribute less to the final output, thereby limiting their impact on model accuracy.

In contrast, Phi-7B exhibits a more noticeable degradation in stall-free mode, particularly at lower thresholds. 
At $\tau=0.60$, the average accuracy drops significantly from 64.6 to 58.6, with large reductions in ARC and MMLU scores. 
Increasing $\tau$ improves performance, with the average recovering to 60.4 at $\tau=0.90$, but a gap relative to the baseline remains. 
The degradation is primarily driven by a subset of benchmarks (notably ARC and MMLU), while others, such as WinoGrande, remain relatively stable.

This behavior is explained by the different routing configuration of Phi models, which use only $k=2$ experts per token. 
In this case, each expert contributes a larger fraction of the final output, making the model more sensitive to mispredictions. 
As a result, even occasional overlap misses can have a more pronounced effect on accuracy compared to Granite models.

\bh{Deployment Guidance:}\label{bh:deployment}
These results show that stall-free execution should be treated as a complementary opt-in mode rather than the default APEX configuration. 
Correctness-preserving APEX always preserves the original routing semantics and is therefore preferable for accuracy-critical deployments. 
Stall-free mode is most appropriate when the application can tolerate small accuracy changes in exchange for additional performance gains. 
Empirically, we observe that models with larger active expert counts, such as Granite-1B/3B with ($k=8$) and DeepSeek-16B with ($k=6$) and two shared experts, are more robust because each missed expert contributes a smaller fraction of the final weighted output.
In contrast, low-top-$k$ models such as Phi-7B ($k=2$) are more sensitive to substitutions; therefore, the correctness-preserving mode is the recommended deployment choice, treating stall-free execution for an optional performance improvement.

\begin{table}[t]
\centering
\caption{Application-level accuracy of APEX in stall-free \textbf{(APEX$_{SF}$)} mode compared to correctness-preserving execution (Base).}
\label{tab:application_acc}
\setlength{\tabcolsep}{3pt}
\resizebox{\linewidth}{!}{%
\begin{tabular}{llccc}
\toprule
Model & Config &\hspace{-1mm}PPL$\downarrow$\hspace{-1mm}&\hspace{-1mm} ARC/MMLU/WG/TQA $\uparrow$\hspace{-1mm}&Avg $\uparrow$\hspace{-1mm}\\
\midrule
\multirow{4}{*}{1B}
 & Base        & \textbf{7.88} & 40.2 / 28.5 / 60.3 / 44.1 & \textbf{43.3} \\
 & APEX$_{SF}$ ($\tau=0.90$) & \underline{7.99} & 39.4 / 28.1 / 60.1 / 43.4 & \underline{42.8} \\
 & APEX$_{SF}$ ($\tau=0.75$) & 8.06 & 39.3 / 28.0 / 60.0 / 43.5 & 42.7 \\
 & APEX$_{SF}$ ($\tau=0.60$) & 8.04 & 39.0 / 28.0 / 60.0 / 43.5 & 42.6 \\
\midrule
\multirow{4}{*}{3B}
 & Base        & \textbf{6.79} & 52.7 / 49.4 / 67.9 / 45.9 & \textbf{54.0} \\
 & APEX$_{SF}$ ($\tau=0.90$) & \underline{6.83} & 52.6 / 49.0 / 67.4 / 44.0 & \underline{53.2} \\
 & APEX$_{SF}$ ($\tau=0.75$) & 6.87 & 52.5 / 48.7 / 67.4 / 43.9 & 53.1 \\
 & APEX$_{SF}$ ($\tau=0.60$) & 6.91 & 52.1 / 48.6 / 66.8 / 43.9 & 52.9 \\
\midrule
\multirow{4}{*}{7B}
 & Base        & \textbf{6.27} & 63.4 / 70.6 / 76.0 / 48.4 & \textbf{64.6} \\
 & APEX$_{SF}$ ($\tau=0.90$) & \underline{7.47} & 57.9 / 62.7 / 74.7 / 46.2 & \underline{60.4} \\
 & APEX$_{SF}$ ($\tau=0.75$) & 7.49 & 57.5 / 62.4 / 74.5 / 45.8 & 60.1 \\
 & APEX$_{SF}$ ($\tau=0.60$) & 7.52 & 51.8 / 62.6 / 74.6 / 45.6 & 58.6 \\
\midrule
\multirow{4}{*}{16B}
 & Base        & \textbf{7.02} & 52.6 / 48.3 / 70.5 / 34.8 & \textbf{51.6} \\
 & APEX$_{SF}$ ($\tau=0.90$) & \underline{7.10} & 52.5 / 46.7 / 70.2 / 33.6 & \underline{50.8} \\
 & APEX$_{SF}$ ($\tau=0.75$) & 7.11 & 52.1 / 46.3 / 70.2 / 33.5 & 50.5 \\
 & APEX$_{SF}$ ($\tau=0.60$) & 7.13 & 52.3 / 45.7 / 66.4 / 31.8 & 49.0 \\

\bottomrule
\end{tabular}
}
\vspace{-5mm}
\end{table}

\subsection{Performance and Energy Analysis}\label{ssec:perf}
This section evaluates the system-level benefits of APEX in terms of latency, energy, and combined EDP efficiency.

\bh{Latency Analysis:}
Figure~\ref{fig:perf_energy_latency}(a) presents the average per-token latency across context lengths for the Granite-3B model as a representative case; other models exhibit similar trends, with combined EDP comparisons provided next. 
At a context length of 512, APEX correctness-preserving mode achieves 11.41~ms average per-token latency. 
This is 42\% lower than no prefetching (19.77 ms) and 26\% lower than ProMoE (15.39 ms).
Similarly, APEX reduces the average latency by 20\% and 40\% relative to static $(k+4)$ and static $(k+8)$ prefetching techniques. 
Longer contexts continue to exhibit the same trend. 
APEX achieves 41\% and 39\% lower latency than no prefetching at 1024 and 2048 token contexts, 
outperforming ProMoE by 24\% and 20\% lower latency. 

Figure~\ref{fig:perf_energy_latency} (a) also show that \emph{static overfetching performs poorly}. 
When too many experts are statically prefetched, expert loading itself begins to dominate and can exceed the attention window available for hiding it, turning prefetch traffic into the new bottleneck.
Finally, we observe that APEX's stall-free mode provides 2.0\%--2.8\% lower latency because it does not stall to bring in missing experts. The latency difference is small since the overlap accuracy is high (i.e., stalls happen rarely).

\bh{Energy Consumption Analysis:}
Figure~\ref{fig:perf_energy_latency}(b) plots the energy consumption results, which reinforce the superiority of APEX over the baselines.
At context length 512, APEX correctness-preserving mode consumes 287.3~mJ energy, which is 9.5\% lower than no prefetching and 5.8\% lower than ProMoE.
We observe that static prefetching leads to the largest energy consumption (9.9\% and 21.8\% more than APEX) due to over-provisioning. 
Results with longer context lengths confirm that both modes of APEX consistently achieve the lowest energy consumption. 
The stall-free mode has again a slightly ($\approx$1\%) lower energy by avoiding additional (rare) prefetches. 
This small gap indicates that correctness-preserving mode already removes nearly all of the exploitable stall overhead, consistent with the very high overlap accuracy reported earlier.

Notably, the static $(k+8)$ is consistently the most energy-consuming policy, as it attempts to avoid misses more aggressively. 
This highlights an important point: overfetching does not come for free. 
Excessive expert transfers increase memory and interconnect activity and can outweigh any stall reduction. 
To further clarify the energy trade-off, Table~\ref{tab:energy} reports an energy breakdown for Granite-3B at 1024-token context. 
APEX does not reduce transfer energy; off-chip I/O energy increases from 48 mJ to 56 mJ due to additional prefetched experts. 
However, by overlapping these transfers with attention computation, APEX reduces exposed stall time and lowers idle/leakage energy from 57 mJ to 8 mJ. 
This reduction more than offsets the added I/O energy, reducing total energy from 340 mJ to 299 mJ. 
In contrast, static (k+8) prefetching overfetches aggressively, increasing I/O energy to 97 mJ and total energy to 377 mJ.
APEX, by comparison, remains close to the Oracle behavior by allocating the additional-prefetch budget needed for each token and layer, which is why it achieves the best combined latency and energy efficiency.

\begin{table}[b]
\centering
\caption{Energy breakdown for Granite-3B at 1024-token context.}
\label{tab:energy}
\scriptsize
\begin{tabular}{lccc}
\toprule
\textbf{Component} & \textbf{No Prefetch} & \textbf{APEX} & \textbf{Static $(k+8)$} \\
\midrule
On-chip active memory & 233 mJ & 234 mJ & 234 mJ\\
Off-chip I/O transfer & 48 mJ & 56 mJ $(+8)$ & 97 mJ\\
Idle / leakage & 57 mJ & 8 mJ $(-49)$ & 44 mJ\\
\midrule
Total energy & 340 mJ & 299 mJ & 377 mJ\\
\bottomrule
\end{tabular}
\end{table}

\begin{figure}[t]
    \centering
    \includegraphics[width=1\linewidth]{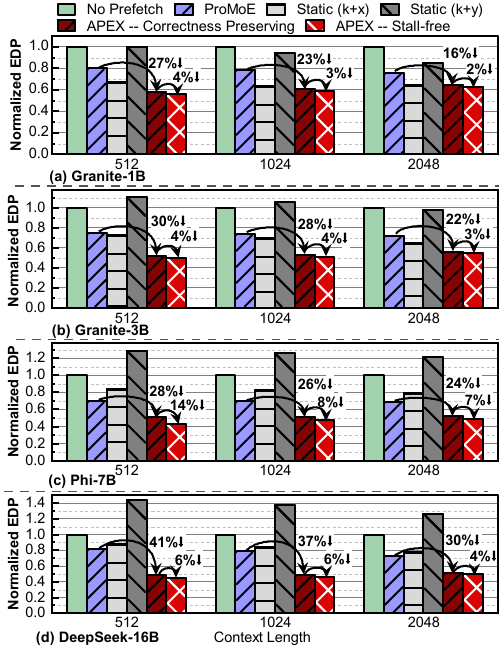}
    \caption{Normalized EDP all models. For Granite models, $k=8$, $x=4$, and $y=8$; for Phi-7B, $k=2$, $x=1$, and $y=2$ and for DeepSeek-16B, $k=6$, $x=4$ and $y=8$. APEX variants consistently achieve the lowest EDP across all context lengths, demonstrating superior hardware efficiency.}
    \label{fig:edp}
    \vspace{-3mm}
\end{figure}

\bh{End-to-End Efficiency (EDP):}
Figure~\ref{fig:edp} shows that APEX consistently achieves the lowest normalized EDP. 
For Granite-1B, APEX correctness-preserving mode lowers the EDP by 36\%--42\% over no prefetching, and 16\%--27\% over ProMoE, across all context lengths.  
The stall-free mode provides an additional of 4\%, 3\%, and 2\% gain over correctness-preserving at 512, 1024, and 2048 tokens, again indicating that residual misses are already rare.
Granite-3B follows the same trend, with APEX achieving the lowest EDP, with 22\%--30\% lower EDP than ProMoE across all context lengths. 
For both models, the static $(k+8)$  consistently has the largest EDP, aligned with the latency and energy analysis, as it attempts to avoid misses more aggressively. 

Both modes of APEX continue to achieve the best EDP compared to baselines for the Phi-7B model, as shown in the third subplot in Figure~\ref{fig:edp}. 
Specifically, the correctness-preserving mode achieves 48\%--49\% lower EDP than no prefetching, and 24\%--28\% lower EDP than ProMoE.  
Unlike the Granite models, the Phi-7B model results in a larger difference between the correctness-preserving and stall-free modes of APEX. 
It lowers the EDP by 14\%, 8\%, and 7\% compared to correctness-preserving for different context lengths.
These results are consistent with Phi’s routing structure and accuracy analysis in Section~\ref{ssec:accuracy}. Each token activates only $k=2$ experts, and missing even one expert has a larger effect than in Granite models, where each token aggregates across $k=8$ experts. 
Thus, eliminating the final residual stalls yields a larger benefit for Phi, at the cost of a larger accuracy degradation.

DeepSeek-16B further confirms the same trend at a larger scale as shown in Figure~\ref{fig:edp} (D). APEX reduces EDP by 30--41\% over ProMoE, while stall-free execution adds a further 4--6\% improvement over correctness-preserving mode. 
This smaller stall-free gap compared to Phi-7B is consistent with DeepSeek's routing structure, where each token uses $k=6$ routed experts plus two shared experts, making occasional routed-expert substitutions less disruptive.

\begin{table}[b]
\vspace{-3mm}
\caption{Overhead analysis of the APEX prefetch router.}
\label{tab:overhead}
\resizebox{\linewidth}{!}{%
\begin{tabular}{@{}lccc@{}}
\toprule
\textbf{Model} & \textbf{Additional Parameters} & \textbf{\% of model weights} & \textbf{\% of Performance} \\ \midrule
\textbf{Granite-1B} & 0.79M & 0.059\% & 0.051\% \\
\textbf{Granite-3B} & 1.97M & 0.060\% & 0.046\% \\
\textbf{Phi-7B} & 2.10M & 0.027\% & 0.009\% \\
\textbf{DeepSeek-16B} & 34.11M & 0.022\% & 0.036\% \\ \bottomrule
\end{tabular}
}
\end{table}

\vspace{-2mm}
\subsection{Overhead Analysis}\label{ssec:overhead}
The additional parameters required by APEX comprise the lightweight prefetch router and the CDF model. 
For example, in Granite-1B, each layer includes a $1024\times32$ prefetch router and $2\times32$ CDF parameters, resulting in a total of 0.79M parameters across 24 layers.  
This is only 0.059\% of the 1.3B model weights. 
The overhead remains similarly negligible for the larger models, accounting for just 0.060\%, 0.027\% and 0.022\% of total parameters for Granite-3B, Phi-7B and DeepSeek-16B, respectively, as shown in Table~\ref{tab:overhead}.

On our target edge platform (described in Section~\ref{ssec:hw}), APEX adds only 0.051\%, 0.046\%, 0.009\% and 0.036\% performance overhead for Granite-1B, Granite-3B, Phi-7B, and DeepSeek-16B, respectively.
Since both the added weights and computation are negligible relative to the base model, this overhead is expected to remain insignificant on other platforms as well.

Overall, these results show that APEX is an extremely low-overhead approach. 
It adds only a lightweight auxiliary predictor on top of the base MoE model, while delivering substantial improvements in latency and energy efficiency.

\subsection{Ablation Study} \label{ssec:ablation}

\begin{figure}[b]
    \centering
    \includegraphics[width=1\linewidth]{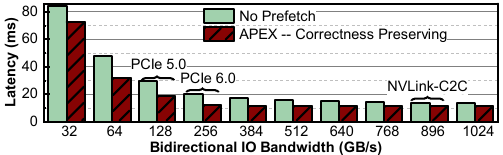}
    \caption{Per-token latency versus bidirectional I/O bandwidth for Granite-3B (1024 context length).} 
    \label{fig:ablation}
\end{figure}

\bh{Sensitivity to I/O Bandwidth:}\label{bh:bw_ablation}
We evaluate APEX across different off-chip I/O bandwidths using Granite-3B as a representative model at a context length of 1024, while keeping all other parameters fixed to isolate the impact of bandwidth on prefetch effectiveness.
The sweep ranges from 32 to 1024 GB/s. 
The low-bandwidth points (32 and 64 GB/s) represent constrained mobile/embedded-class settings where expert transfers receive only a fraction of total DRAM bandwidth due to sharing with CPU/GPU/NPU activity, KV-cache traffic, and memory-controller contention. 
The 128 and 256 GB/s points correspond to PCIe 5.0 $\times$16 and PCIe 6.0 $\times$16 bandwidths, respectively.
As expected, increasing bandwidth reduces latency, even for the no-prefetch baseline, since expert transfers become less expensive at higher bandwidth.
Figure~\ref{fig:ablation} shows that APEX consistently decreases the latency by 14\%--42\% across the entire bandwidth range by minimizing the number of exposed expert-loading stalls. 
At very low bandwidths, the benefit is limited by the fact that only part of the expert transfer can be hidden behind the attention window, so the unhidden portion remains exposed as stall time. 
However, APEX still consistently improves latency across the full range by overlapping the transferable portion of expert loading with attention and by avoiding the unnecessary traffic of fixed overfetching.
At the same time, APEX remains beneficial even at high-bandwidth settings approaching NVLink-C2C class interconnects, showing that adaptive prefetching continues to improve latency even when I/O is less constrained. 
Overall, these results highlight that APEX is robust across a wide range of deployment scenarios, with particularly strong benefits in the bandwidth-limited edge setting. 

\begin{figure}[t]
    \centering
    \includegraphics[width=1\linewidth]{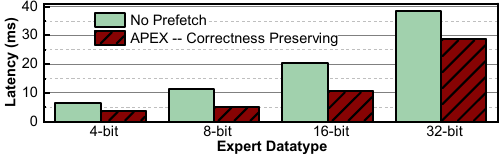}
    \caption{Per-token latency sensitivity to expert-weight datatype for Granite-3B (1024 context length).}
    \label{fig:ablation_datatype}
    \vspace{-5mm}
\end{figure}

\bh{Sensitivity to Expert Weight Precision:}\label{bh:data_ablation} 
Since many edge deployments use quantized weights, we further evaluate APEX across different expert-weight data types on Granite-3B (1024 context). 
This study scales the expert transfer size and compute cost for 4-, 8-, 16-, and 32-bit expert weights while keeping the routing behavior fixed. 
Therefore, it isolates the system-level impact of datatype and does not quantify quantization-induced accuracy loss. 
As shown in Figure~\ref{fig:ablation_datatype}, reducing precision lowers absolute latency for both No Prefetch and APEX because each expert transfer becomes smaller. 
However, APEX continues to reduce latency across all datatypes by overlapping expert movement with attention computation. 
The benefit is especially clear when expert transfers remain exposed on the critical path, while at very low precision, the execution is dominated by attention. 
These results show that APEX complements quantization. Quantization reduces the amount of data moved~\cite{sun2026lexi}, whereas APEX reduces the portion of that movement exposed.

\vspace{-2mm}
\section{Conclusion} \label{sec:conclusion}

Edge MoE inference is fundamentally bottlenecked by expert I/O, where irregular and off-chip expert accesses stall execution and waste energy. 
In this work, we presented APEX, an adaptive expert prefetching framework that transforms expert loading into a predictive, confidence-driven decision. 
By introducing a lightweight prefetch router and a learned CDF-based mechanism to dynamically select the minimal top-$(k+\hat{\delta}(x))$ prefetch budget, APEX achieves near-oracle expert coverage while avoiding the inefficiencies of static overfetching.
This adaptive approach delivers substantial system-level benefits. 
At higher confidence thresholds, APEX achieves $>$99\% overlap accuracy and effectively eliminates expert-loading stalls while preserving application-level accuracy. 
Across multiple models and settings, it reduces latency and improves energy-delay product (EDP) by up to 41\% compared to state-of-the-art baselines, with negligible model and runtime overhead. 
Overall, APEX demonstrates that adaptive, confidence-aware prefetching is key to unlocking efficient MoE inference on edge systems, bridging the gap between predictive accuracy and system performance.

\noindent \textit{Disclosure}: Dr. Ogras is affiliated with Samsung Austin Research \& Development Center and Advanced Computing Lab (SARC/ACL). This relationship has been approved under applicable outside activities policies.

\bibliographystyle{IEEEtran}
\bibliography{refs/refs_codex}

@article{zheng2025review,
  author = {Zheng, Yue and others},
  title = {A Review on Edge Large Language Models: Design, Execution, and Applications},
  journal = {ACM Computing Survey},
  volume = {57},
  number = {8},
  pages = {1--35},
  year = {2025}
}

@inproceedings{park2024thermal,
  author = {Park, Jaehyun and others},
  title = {Thermal Modeling and Management Challenges in Heterogeneous Integration: {2.5D} Chiplet Platforms and Beyond},
  booktitle = {Proc. {IEEE} VLSI Test Symp. ({VTS})},
  pages = {1--4},
  year = {2024}
}

@article{pfromm2025mfit,
  author    = {Lukas Pfromm and others},
  title     = {{MFIT}: Multi-{Fidelity} {Thermal} {Modeling} for 2.5D and 3D {Multi}-{Chiplet} {Architectures}},
  journal   = {ACM Transactions on Design Automation of Electronic Systems},
  year      = {2025},
  doi       = {10.1145/3765905}
}

@article{kanani2025thermos,
  author    = {Alish Kanani and others},
  title     = {{THERMOS}: Thermally-Aware {Multi}-{Objective} {Scheduling} of {AI} {Workloads} on {Heterogeneous} {Multi}-{Chiplet} {PIM} {Architectures}},
  journal={ACM Transactions on Embedded Computing Systems},
  volume={24},
  number={5s},
  pages={1--26},
  year={2025}
}

@article{meuser2024revisiting,
  author = {Meuser, Tobias and others},
  title = {Revisiting Edge {AI}: Opportunities and Challenges},
  journal = {IEEE Internet Computing},
  volume = {28},
  number = {4},
  pages = {49--59},
  year = {2024}
}

@inproceedings{he2025hydra,
  author = {He, Siqi and others},
  title = {{Hydra}: Harnessing Expert Popularity for Efficient Mixture-of-Expert Inference on Chiplet Systems},
  booktitle = {Proc. {ACM/IEEE} Design Automation Conf. ({DAC})},
  pages = {1--7},
  year = {2025}
}

@inproceedings{dong2025ubimoe,
  author = {Dong, Jiale and others},
  title = {{UbiMoE}: A Ubiquitous Mixture-of-Experts Vision Transformer Accelerator with Hybrid Computation Pattern on {FPGA}},
  booktitle = {Proc. {IEEE} Int. Symp. Circuits Systems ({ISCAS})},
  pages = {1--5},
  year = {2025}
}

@inproceedings{zou2024fed,
  author = {Zou, Yifei and others},
  title = {{FED-MOE}: Efficient Federated Learning for Mixture-of-Experts Models via Empirical Pruning},
  booktitle = {Proc. Int. Conf. Parallel Distributed Computing: Applications and technologies ({PDCAT})},
  pages = {128--139},
  year = {2024}
}

@misc{cheng2025ermoe,
  author = {Cheng, Anzhe and others},
  title = {{ERMoE}: Eigen-Reparameterized Mixture-of-Experts for Stable Routing and Interpretable Specialization},
  year = {2025},
  eprint = {2511.10971},
  archiveprefix = {arXiv},
  primaryclass = {cs.CV}
}

@article{gholami2024ai,
  author = {Gholami, Amir and others},
  title = {{AI} and the Memory Wall},
  journal = {IEEE Micro},
  volume = {44},
  number = {3},
  pages = {33--39},
  year = {2024}
}

@inproceedings{vaswani2017attention,
  author = {Vaswani, Ashish and others},
  title = {Attention Is All You Need},
  booktitle = {Proc. Adv. Neural Information Processing System ({NeurIPS})},
  volume = {30},
  pages = {5998--6008},
  year = {2017}
}

@article{fedus2022switch,
  author = {Fedus, William and Zoph, Barret and Shazeer, Noam},
  title = {Switch Transformers: Scaling to Trillion Parameter Models with Simple and Efficient Sparsity},
  journal = {J. Machine Learning Research},
  volume = {23},
  number = {120},
  pages = {1--39},
  year = {2022}
}

@article{jiang2024mixtral,
  author = {Jiang, Albert Q. and others},
  title = {Mixtral of Experts},
  year = {2024},
  eprint = {2401.04088},
  archiveprefix = {arXiv},
  primaryclass = {cs.LG}
}

@inproceedings{lepikhin2020gshard,
  author = {Lepikhin, Dmitry and others},
  title = {{GShard}: Scaling Giant Models with Conditional Computation and Automatic Sharding},
  booktitle = {Proc. Int. Conf. Learning Representations ({ICLR})},
  year = {2021}
}

@inproceedings{pope2023efficiently,
  author = {Pope, Reiner and others},
  title = {Efficiently Scaling Transformer Inference},
  booktitle = {Proc. Maching Learning System ({MLSys})},
  volume = {5},
  year = {2023}
}

@inproceedings{shazeer2017outrageously,
  author = {Shazeer, Noam and others},
  title = {Outrageously Large Neural Networks: The Sparsely-Gated Mixture-of-Experts Layer},
  booktitle = {Proc. Int. Conf. Learning Representations ({ICLR})},
  year = {2017}
}

@misc{zoph2022st,
  author = {Zoph, Barret and others},
  title = {{ST-MoE}: Designing Stable and Transferable Sparse Expert Models},
  year = {2022},
  eprint = {2202.08906},
  archiveprefix = {arXiv},
  primaryclass = {cs.LG}
}

@inproceedings{lewis2021base,
  author = {Lewis, Mike and Bhosale, Shruti and Dettmers, Tim and Goyal, Naman and Zettlemoyer, Luke},
  title = {{BASE} Layers: Simplifying Training of Large, Sparse Models},
  booktitle = {Proc. Int. Conf. Machine Learning ({ICML})},
  pages = {6265--6274},
  year = {2021}
}

@inproceedings{dai2024deepseekmoe,
  author = {Dai, Damai and others},
  title = {{DeepSeekMoE}: Towards Ultimate Expert Specialization in Mixture-of-Experts Language Models},
  booktitle = {Proc. Annu. Meet. Assoc. Computational Linguistics ({ACL})},
  pages = {1280--1297},
  year = {2024}
}

@misc{muennighoff2024olmoe,
  author = {Muennighoff, Niklas and others},
  title = {{OLMoE}: Open Mixture-of-Experts Language Models},
  year = {2024},
  eprint = {2409.02060},
  archiveprefix = {arXiv},
  primaryclass = {cs.CL}
}

@misc{he2021fastmoe,
  author = {He, Jiaao and others},
  title = {{FastMoE}: A Fast Mixture-of-Expert Training System},
  year = {2021},
  eprint = {2103.13262},
  archiveprefix = {arXiv},
  primaryclass = {cs.DC}
}

@inproceedings{gale2023megablocks,
  author = {Gale, Trevor and Narayanan, Deepak and Young, Cliff and Zaharia, Matei},
  title = {{MegaBlocks}: Efficient Sparse Training with Mixture-of-Experts},
  booktitle = {Proc. Machine Learning and Systems ({MLSys})},
  volume = {5},
  year = {2023}
}

@misc{NVIDIA_RTX3090,
  author = {{NVIDIA Corporation}},
  title = {{GeForce RTX 3090}},
  year = {2020},
  url = {https://www.nvidia.com/en-us/geforce/graphics-cards/30-series/rtx-3090/},
  note = {accessed 2026-03-11}
}

@misc{NVIDIA_A100,
  author = {{NVIDIA Corporation}},
  title = {{NVIDIA A100 Tensor Core GPU Architecture}},
  year = {2020},
  url = {https://www.nvidia.com/en-us/data-center/a100/},
  note = {accessed 2026-06-01}
}

@misc{nvidia_jetson_thor,
  author       = {{NVIDIA}},
  title        = {{Jetson Thor: Advanced AI for Physical Robotics}},
  year         = {2026},
  howpublished = {\url{https://www.nvidia.com/en-us/autonomous-machines/embedded-systems/jetson-thor/}},
  note         = {Accessed: 2026-06-14}
}

@misc{hailo_10h_accelerator,
  author       = {{Hailo}},
  title        = {{Hailo-10H AI Accelerator: LLM and VLM Generative AI Accelerator}},
  year         = {2024},
  howpublished = {\url{https://hailo.ai/products/ai-accelerators/hailo-10h-ai-accelerator/}},
  note         = {Accessed: 2026-06-14}
}

@misc{granite31_a800m,
  author = {{IBM Research}},
  title = {{Granite-3.1-3B-A800M Base}: A Sparse Mixture-of-Experts Language Model},
  year = {2024},
  url = {https://huggingface.co/ibm-granite/granite-3.1-3b-a800m-base},
  note = {model card, accessed 2026-03-27}
}

@misc{granite31_1b_a400m,
  author = {{IBM Research}},
  title = {{Granite-3.1-1B-A400M Base}: A Sparse Mixture-of-Experts Language Model},
  year = {2024},
  url = {https://huggingface.co/ibm-granite/granite-3.1-1b-a400m-base},
  note = {model card, accessed 2026-03-27}
}

@misc{phi3_technical_report,
  author = {Abdin, Marah and others},
  title = {{Phi-3} Technical Report: A Highly Capable Language Model Locally on Your Phone},
  year = {2024},
  eprint = {2404.14219},
  archiveprefix = {arXiv},
  primaryclass = {cs.CL}
}

@article{liu2024deepseekv2,
  title={Deepseek-v2: A strong, economical, and efficient mixture-of-experts language model},
  author={Liu, Aixin and others},
  journal={arXiv preprint arXiv:2405.04434},
  year={2024}
}

@book{goodfellow2016deep,
  author = {Goodfellow, Ian and Bengio, Yoshua and Courville, Aaron},
  title = {Deep Learning},
  publisher = {MIT Press},
  year = {2016}
}

@article{zhu2025pre,
  title={Pre-Attention Expert Prediction and Prefetching for Mixture-of-Experts Large Language Models},
  author={Zhu, Shien and Bohl, Samuel and Oester, Robin and Alonso, Gustavo},
  journal={arXiv preprint arXiv:2511.10676},
  year={2025}
}

@misc{song2024promoe,
  author = {Song, Xiaoniu and Zhong, Zihang and Chen, Rong and Chen, Haibo},
  title = {{ProMoE}: Fast {MoE}-Based {LLM} Serving Using Proactive Caching},
  year = {2024},
  eprint = {2410.22134},
  archiveprefix = {arXiv},
  primaryclass = {cs.DC}
}

@misc{xue2024moe,
  author = {Xue, Leyang and others},
  title = {{MoE-Infinity}: Efficient {MoE} Inference on Personal Machines with Sparsity-Aware Expert Cache},
  year = {2024},
  eprint = {2401.14361},
  archiveprefix = {arXiv},
  primaryclass = {cs.LG}
}

@misc{fang2025fate,
  author = {Fang, Zhiyuan and others},
  title = {{FATE}: Fast Edge Inference of Mixture-of-Experts Models via Cross-Layer Gate},
  year = {2025},
  eprint = {2502.12224},
  archiveprefix = {arXiv},
  primaryclass = {cs.LG}
}

@inproceedings{zhong2024adapmoe,
  author = {Zhong, Shuzhang and others},
  title = {{AdapMoE}: Adaptive Sensitivity-Based Expert Gating and Management for Efficient {MoE} Inference},
  booktitle = {Proc. {IEEE/ACM} Int. Conf. Computer-Aided Design ({ICCAD})},
  pages = {1--9},
  year = {2024}
}

@misc{tang2024hobbit,
  author = {Tang, Peng and others},
  title = {{HOBBIT}: A Mixed Precision Expert Offloading System for Fast {MoE} Inference},
  year = {2024},
  eprint = {2411.01433},
  archiveprefix = {arXiv},
  primaryclass = {cs.DC}
}

@inproceedings{hwang2024pre,
  author = {Hwang, Ranggi and others},
  title = {Pre-gated {MoE}: An Algorithm-System Co-Design for Fast and Scalable Mixture-of-Expert Inference},
  booktitle = {Proc. Annu. Int. Symp. Computer Architecture ({ISCA})},
  pages = {1018--1031},
  year = {2024}
}

@article{li2026moe,
  author = {Li, Shuhuai and Lin, Jianghao and Ge, Dongdong and Ye, Yinyu},
  title = {{MoE-SpAc}: Efficient {MoE} Inference Based on Speculative Activation Utility in Heterogeneous Edge Scenarios},
  year = {2026},
  eprint = {2603.09983},
  archiveprefix = {arXiv},
  primaryclass = {cs.LG}
}

@article{eliseev2023fast,
  author = {Eliseev, Artyom and Mazur, Denis},
  title = {Fast Inference of Mixture-of-Experts Language Models with Offloading},
  year = {2023},
  eprint = {2312.17238},
  archiveprefix = {arXiv},
  primaryclass = {cs.LG}
}

@article{merity2016wiki,
  author = {Merity, Stephen and Xiong, Caiming and Bradbury, James and Socher, Richard},
  title = {Pointer Sentinel Mixture Models},
  year = {2016},
  eprint = {1609.07843},
  archiveprefix = {arXiv},
  primaryclass = {cs.CL},
  note = {Introduces the {WikiText} language modeling datasets}
}

@article{clark2018think,
  author = {Clark, Peter and others},
  title = {Think You Have Solved Question Answering? Try {ARC}, the {AI2} Reasoning Challenge},
  year = {2018},
  eprint = {1803.05457},
  archiveprefix = {arXiv},
  primaryclass = {cs.AI}
}

@article{hendrycks2020measuring,
  author = {Hendrycks, Dan and others},
  title = {Measuring Massive Multitask Language Understanding},
  year = {2020},
  eprint = {2009.03300},
  archiveprefix = {arXiv},
  primaryclass = {cs.CY}
}

@article{sakaguchi2021winogrande,
  author = {Sakaguchi, Keisuke and Le Bras, Ronan and Bhagavatula, Chandra and Choi, Yejin},
  title = {{WinoGrande}: An Adversarial Winograd Schema Challenge at Scale},
  journal = {Commun. ACM},
  volume = {64},
  number = {9},
  pages = {99--106},
  year = {2021}
}

@inproceedings{lin2022truthfulqa,
  author = {Lin, Stephanie and Hilton, Jacob and Evans, Owain},
  title = {{TruthfulQA}: Measuring How Models Mimic Human Falsehoods},
  booktitle = {Proc. Annu. Meet. Assoc. computational linguistics ({ACL})},
  pages = {3214--3252},
  year = {2022}
}

@inproceedings{reggiani2023flex,
  author = {Reggiani, Enrico and Andri, Renzo and Cavigelli, Lukas},
  title = {{Flex-SFU}: Accelerating {DNN} Activation Functions by Non-Uniform Piecewise Approximation},
  booktitle = {Proc. {ACM/IEEE} Design Automation Conference ({DAC})},
  pages = {1--6},
  year = {2023}
}

@article{park2022192,
  author = {Park, Myeong-Jae and others},
  title = {A 192-Gb 12-High 896-GB/s {HBM3 DRAM} with a {TSV} Auto-Calibration Scheme and Machine-Learning-Based Layout Optimization},
  journal = {IEEE J. Solid-State Circuits},
  volume = {58},
  number = {1},
  pages = {256--269},
  year = {2023}
}

@misc{pcie6_spec,
  author = {{PCI-SIG}},
  title = {{PCI Express Base Specification Revision 6.0}},
  year = {2022},
  url = {https://pcisig.com/specifications/pciexpress/base6},
  note = {{PCI-SIG} specification}
}

@inproceedings{ghose2018dram,
  author = {Ghose, Saugata and others},
  title = {Understanding {DRAM} Power Consumption: Experimental Characterization and Analysis},
  booktitle = {Proc. {ACM SIGMETRICS}},
  year = {2018}
}

@inproceedings{kwon2023efficient,
  author = {Kwon, Woosuk and others},
  title = {Efficient Memory Management for Large Language Model Serving with {PagedAttention}},
  booktitle = {Proc. ACM Symp. operating systems principles ({SOSP})},
  pages = {611--626},
  year = {2023}
}

@inproceedings{aminabadi2022deepspeed,
  author = {Aminabadi, Reza Yazdani and others},
  title = {{DeepSpeed-Inference}: Enabling Efficient Inference of Transformer Models at Unprecedented Scale},
  booktitle = {Proc. Int. Conf. Performance Computing, Networking, Storage and Analysis ({SC})},
  year = {2022}
}

@misc{primetime,
  author = {{Synopsys Inc.}},
  title = {{Synopsys PrimeTime}},
  year = {2023},
  url = {https://www.synopsys.com/implementation-and-signoff/signoff/primetime.html},
  note = {static timing and power analysis tool}
}

@article{balasubramonian2017cacti,
  author = {Balasubramonian, Rajeev and Kahng, Andrew B. and Muralimanohar, Naveen and Shafiee, Ali and Srinivas, Vaishnav},
  title = {{CACTI} 7: New Tools for Interconnect Exploration in Innovative Off-Chip Memories},
  journal = {ACM Trans. Architecture and Code Optimization (TACO)},
  volume = {14},
  number = {2},
  pages = {1--25},
  year = {2017}
}

@article{luo2023ramulator,
  author = {Luo, Haocong and others},
  title = {{Ramulator 2.0}: A Modern, Modular, and Extensible {DRAM} Simulator},
  journal = {IEEE Computer Architecture Letters},
  volume = {23},
  number = {1},
  pages = {112--116},
  year = {2024}
}

@article{sharma2022universal,
  author = {Sharma, Debendra Das and Pasdast, Gerald and Qian, Zhiguo and Aygun, Kemal},
  title = {Universal Chiplet Interconnect Express ({UCIe}): An Open Industry Standard for Innovations with Chiplets at Package Level},
  journal = {IEEE Trans. Components, Packaging and Manufacturing Technology},
  volume = {12},
  number = {9},
  pages = {1423--1431},
  year = {2022}
}

@article{pfromm2025chipsim,
  author = {Pfromm, Lukas and others},
  title = {{CHIPSIM}: A Co-Simulation Framework for Deep Learning on Chiplet-Based Systems},
  journal = {IEEE Open J. Solid-State Circuits Society},
  volume={5},
  pages={410-423},
  year = {2025}
}

@article{shan2022architecture,
  author = {Shan, Guangbao and Zheng, Yanwen and Xing, Chaoyang and Chen, Dongdong and Li, Guoliang and Yang, Yintang},
  title = {Architecture of Computing System Based on Chiplet},
  journal = {Micromachines},
  volume = {13},
  number = {2},
  pages = {205},
  year = {2022}
}

@inproceedings{alish2026duet,
  author       = {Alish Kanani and others},
  title        = {{DUET}: {Disaggregated} {Hybrid} {Mamba}-{Transformer} {LLMs} with {Prefill} and {Decode}-{Specific} {Packages}},
  booktitle = {Proc. {ACM/IEEE} Design Automation Conf. ({DAC})},
  pages = {1--7},
  year = {2026}
}

@inproceedings{sun2026lexi,
  author       = {Miao Sun and Alish Kanani and Kaushik Shroff and Umit Ogras},
  title        = {{LEXI}: {Lossless} {Exponent} {Coding} for {Efficient} {Inter}-{Chiplet} {Communication} in {Hybrid} {LLMs}},
  booktitle = {Proc. {ACM/IEEE} Design Automation Conf. ({DAC})},
  pages = {1--7},
  year = {2026}}

\end{document}